\documentclass[10pt,final,journal]{IEEEtran} 

\IEEEoverridecommandlockouts
\usepackage[noadjust]{cite}
\usepackage{amsmath,amssymb,amsfonts}
\usepackage{algorithmic}
\usepackage{graphicx,booktabs}
\usepackage{textcomp}
\usepackage[caption=false,font=footnotesize]{subfig}

\def\BibTeX{{\rm B\kern-.05em{\sc i\kern-.025em b}\kern-.08em
    T\kern-.1667em\lower.7ex\hbox{E}\kern-.125emX}}

\newcommand{\expval}[1]{\operatorname{E}[#1]}
\newcommand{\variance}{\operatorname{var}}

\begin{document}

\title{Receiver Operating Characteristics for a Prototype Quantum Two-Mode Squeezing Radar}

\author{David Luong, C. W. Sandbo Chang, A. M. Vadiraj, Anthony Damini,~\IEEEmembership{Senior Member, IEEE},\\
	C. M. Wilson, and Bhashyam Balaji,~\IEEEmembership{Senior Member, IEEE}
	\thanks{D. Luong, A. Damini, and B. Balaji are with the Radar Sensing and Exploitation Section, Defence Research and Development Canada, 3701 Carling Ave, Ottawa, Ontario, Canada. Email: \{david.luong, anthony.damini, bhashyam.balaji\}@drdc-rddc.gc.ca}% <-this % stops a space
	\thanks{C. W. S. Chang, A. M. Vadiraj, and C. M. Wilson are with the Institute of Quantum Computing, University of Waterloo, 200 University Ave W, Waterloo, Ontario, Canada. Email: \{sandbo.chang, vmananthapadmanabharao, chris.wilson\}@uwaterloo.ca}% <-this % stops a space
}

\markboth{IEEE Transactions on Aerospace and Electronic Systems}{IEEE Transactions on Aerospace and Electronic Systems}

\maketitle

\begin{abstract}
	We have built and evaluated a prototype quantum radar, which we call a \emph{quantum two-mode squeezing radar} (QTMS radar), in the laboratory. It operates solely at microwave frequencies; there is no downconversion from optical frequencies. Because the signal generation process relies on quantum mechanical principles, the system is considered to contain a quantum-enhanced radar transmitter. This transmitter generates a pair of entangled microwave signals and transmits one of them through free space, where the signal is measured using a simple and rudimentary receiver.
	
	At the heart of the transmitter is a device called a Josephson parametric amplifier (JPA), which generates a pair of entangled signals called two-mode squeezed vacuum (TMSV) at 6.1445 GHz and 7.5376 GHz. These are then sent through a chain of amplifiers. The 7.5376 GHz beam passes through 0.5 m of free space; the 6.1445 GHz signal is measured directly after amplification. The two measurement results are correlated in order to distinguish signal from noise.
	
	We compare our QTMS radar to a classical radar setup using conventional components, which we call a \emph{two-mode noise radar} (TMN radar), and find that there is a significant gain when both systems broadcast signals at \textminus82 dBm. This is shown via a comparison of receiver operator characteristic (ROC) curves. In particular, we find that the quantum radar requires \emph{8 times fewer integrated samples} compared to its classical counterpart to achieve the same performance.	
\end{abstract}


\section{Introduction}
\label{sec:intro}

At heart, radars are simple: they transmit radio waves at a target and measure the echos to infer the presence of a target. This simple task, however, is complicated by all sorts of confounding factors such as clutter, jammers, and noise. Numerous strategies, some dating back to the earliest days of radar, have been devised to overcome such interference. One of the newest contenders is \emph{quantum radar}.

Any type of radar that exploits features unique to quantum mechanics (such as entanglement, which is absent in classical physics) to enhance detection ability can be called a quantum radar. There exist a number of theoretical proposals for various types of quantum radars, such as interferometric quantum radar \cite{dowling2008n00n,lanzagorta2011quantum} and quantum illumination radar \cite{lloyd2008qi,tan2008quantum,zhang2014qiphoton,zhang2015entanglement,liu2017discrete}. The former shows improved parameter estimation at high SNR compared to conventional radars, but perform worse at low SNR. The latter is one of the most promising approaches because quantum information theory suggests that such a radar would outperform an ``optimum'' classical radar in the low-SNR regime. Quantum illumination makes use of a phenomenon called \emph{entanglement}, which is in effect a strong type of correlation, to distinguish between signal and noise. The standard quantum illumination protocol can be summarized as follows: generate two entangled pulses of light, send one of them toward a target, and perform a simultaneous measurement on the echo and the other entangled pulse. The measurement result will be different depending on whether the received signal was a true echo or simply uncorrelated noise.

It is difficult to perform a joint, simultaneous measurement on the two entangled pulses unless the distance to the target is already known. Therefore, the original quantum illumination protocol described in \cite{lloyd2008qi} is impractical for radar purposes. 

In this paper, we describe a prototype quantum radar which is inspired by quantum illumination, but which requires only \emph{independent} measurement of the two pulses. Moreover, the measurements are simply of in-phase and quadrature voltages, which are easily performed using off-the-shelf equipment.

Our quantum radar prototype, which we call a \emph{quantum two-mode squeezing radar} (QTMS radar), generates a pair of entangled microwave signals at 6.1445 GHz and 7.5376 GHz. The particular type of entanglement we use is called \emph{two-mode squeezed vacuum} (TMSV), which explains the name we have chosen for our prototype. These entangled signals pass through a chain of amplifiers and are split into two paths. One of them ends at a digitizer which performs measurements at 6.1445 GHz; the other leads to a horn antenna. The transmitted signal is then received at another horn antenna and measured at 7.5376 GHz. With matched filtering and a suitable detector, we can infer the presence or absence of a target and effectively increase the SNR.

In our system, the only ``quantum'' component is the signal generator, so we may be said to have developed a quantum-enhanced radar transmitter. This signal generator is called a Josephson parametric amplifier (JPA). As part of our analysis, we compare the JPA signals with a more conventional system that replaces the JPA with a source of classically correlated (but \emph{not} entangled) signals. We find that there is a definite quantum enhancement with the QTMS radar over this classical radar, which we call a two-mode noise radar (TMN radar). In our setup, the classical system requires an integration time \emph{eight times longer} than that of the quantum system in order to achieve similar performance. This was quantified using receiver operator characteristic (ROC) curves.

Our system is similar to a quantum illumination radar in that it exploits entanglement, but as mentioned above, it does not perform a joint measurement of the two beams. It also shows similarities to noise radars, which have been studied in the radar literature. However, it does not operate on exactly the same principles as a conventional noise radar, either. The entangled beams are not identical---they are not even at the same wavelength---and the covariance structure between them is different from that expected in a noise radar. For these reasons, we have used the name \emph{quantum two-mode squeezing radar}.

An abbreviated, preliminary version of this paper was submitted to the 2019 IEEE Radar Conference. In the following, we have added details and performed a more thorough analysis. The present paper and the 2019 IEEE Radar Conference paper build on work previously done by us and described in \cite{luong2018mqr}, in which both signals were measured directly without passing through free space. Now that we have installed antennas, we may justifiably claim that our device is a quantum radar. A separate analysis of our work from the quantum physics point of view, describing its novelty vis-\`{a}-vis conventional quantum illumination, is presented in \cite{chang2018quantum}.

Note that in \cite{chang2018quantum}, we called our prototype a \emph{quantum-enhanced noise radar}. This is because it does indeed have strong similarities to noise radar. However, in the present paper we dive deeper into the radar signal processing aspects. In this context the difference between conventional noise radars and our prototype are important, so we adopt more distinct name \emph{quantum two-mode squeezing radar}.

\subsection{Claims of This Paper}

To the best of our knowledge, no experimental quantum radar---in the strict sense of the term---has ever been described in the scientific literature. All previous ``quantum radar'' implementations operate at optical wavelengths and are, strictly speaking, quantum \emph{lidars} \cite{lopaeva2013qi,zhang2015entanglement,england2018quantum,balaji2018qi}. Although there are claims in the popular press that quantum radars have already been built \cite{balaji2018snake}, \emph{we are the first to describe an experimental implementation of a quantum radar protocol in a scientific publication}. (This should be understood in the sense that our prototype uses a quantum signal generator. The signals exiting the horns are not entangled, but they remain strongly correlated.)

In this paper, we describe how:
\begin{itemize}
	\item we have built an experimental apparatus which generates an entangled microwave signal. It produces the signal directly in the microwave, without downconversion from optical frequencies.
	\item we require cryogenic temperatures to generate the signal, but after amplification the signal exits the refrigerator. Therefore, we do not require targets to be inside the refrigerator in order to detect it. We broadcast, receive, and measure the signal using conventional horn antennas and digitizers.
	\item our protocol does not require joint measurement of the two entangled beams. Thus we do not need quantum memories or delay lines.
	\item our QTMS radar prototype shows substantial improvements over a similar radar system using non-entangled signals, as quantified using receiver operator characteristic (ROC) curves. This is true even though the amplification process breaks the entanglement.
	\item quantum radar is not infeasible. It is worth looking into because the possible gains are large.
\end{itemize}
\emph{We do not claim that our prototype is in any way ideal.} On the contrary, it is merely an adaptation of existing experimental apparatus. We did not invent the JPA, the source of the entangled signal, either; we only noticed that it could be used in a radar context. Improvements are possible in many aspects of the experiment. There is much work to be done on the radar engineering, signal processing, and quantum physics aspects of this prototype.

\section{The Basic Idea}
\label{sec:abstract_setup}

In the abstract, the operating principle behind our prototype QTMS radar is quite simple. It generates two correlated signals, transmits one of them, and exploits the correlations to distinguish the transmitted signal from external noise. To put it another way, we perform matched filtering between the two generated signals after one of them is transmitted toward a target receive antenna.

We have built two simple radar prototypes based on this principle: a QTMS radar and a ``classical'' radar which attempts to approximate the output of the quantum radar using conventional methods. In both cases, the two correlated signals consist of noise: pseudo-random noise in the classical case, true random noise in the quantum case.

\begin{figure}[h]
	\centerline{\includegraphics[width=.7\columnwidth]{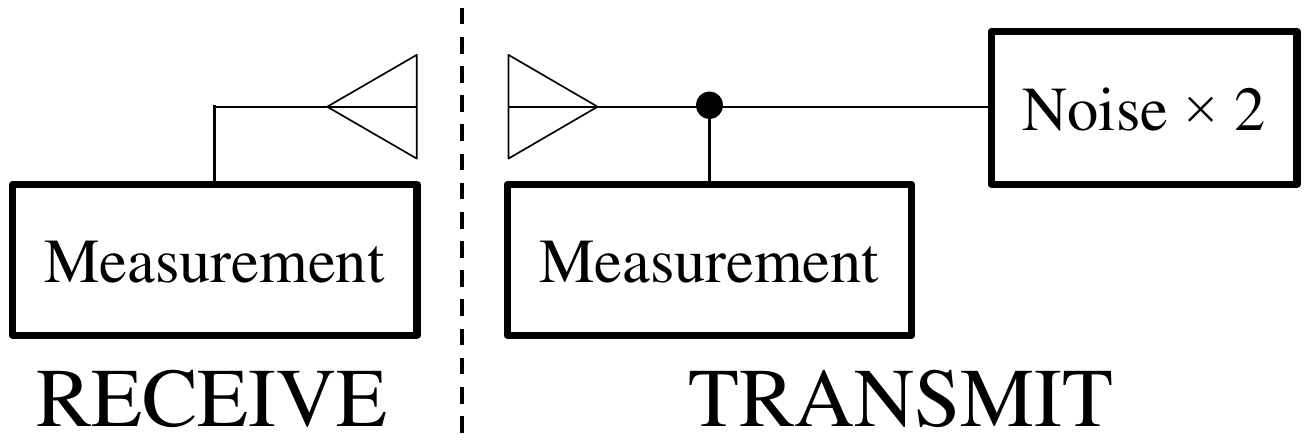}}
	\caption{Block diagram illustrating the basic idea behind our QTMS radar.}
	\label{fig:block_diagram_abstract}
\end{figure}

Figure \ref{fig:block_diagram_abstract} is a block diagram depicting the basic setup of both prototypes. We may summarize the operation of the radars as follows:
\begin{enumerate}
	\item Produce two correlated noise signals.
	\item Measure one signal immediately. Retain a record of the measured in-phase and quadrature voltages. Transmit the other signal through free space.
	\item Receive and measure the transmitted signal.
	\item Declare a detection if the detector output, based on the matched filter output generated by suitably correlating the received and recorded signals, exceeds a certain threshold.
\end{enumerate}
In fact, the only difference between our quantum and classical setups is in the first step---that is, in the generation of the correlated noise signals. Apart from this, we treat the two setups identically: we measure in-phase and quadrature voltages in both cases and perform the same data analysis using the same computer code.

We note that the basic concept described here is quite similar to \emph{noise radar} as described in the radar literature (e.g. in \cite{narayanan2016noise,dawood2001roc}). Indeed, they are similar enough that in some of the analysis to follow, we have made use of certain noise radar results. However, our scheme is a slight generalization in the sense that we do not require the two generated noise signals to be the \emph{same}; they need only be \emph{correlated}. The correlations should be as strong as possible to maximize the output of the matched filter. We will see that the quantum setup achieves better results than the classical one because it generates stronger correlations.

In the following, we denote the in-phase and quadrature components of the two generated signals by $I_1$, $Q_1$, $I_2$, and $Q_2$. The subscripts 1 and 2 serve to distinguish the two signals. We will assume that signal 1 is transmitted through free space while signal 2 is measured immediately. To avoid a proliferation of subscripts, we will use the same variables to refer to the signals that reach the measurement apparatus. Context will serve to distinguish the two cases.

Though it is common to combine in-phase and quadrature voltages into a single complex number, for example $z_1 = I_1 + jQ_1$, we have chosen not to do so in this paper. This is to avoid confusion because real-valued voltages are more natural in quantum physics.

\subsection{Covariance Matrices in a Single-Channel System}

This paper frequently discusses covariance matrices formed from measured voltages---that is, matrices of the form $\expval{xx^T}$ where $x = [I_1, Q_1, I_2, Q_2]^T$. It is important to understand that these matrices have no relation whatsoever to the covariance matrices that arise in array processing. Our setup is a single-channel system. The voltages that go into this matrix are not outputs of a matched filter; they are the inputs. The recorded voltages $I_2$ and $Q_2$ are used as a reference for matched filtering of the received signals $I_1$ and $I_2$.

Note, also, that signal 1 and signal 2 are generally measured at different times. The delay between the two is related to the free-space path length of the transmitted signal. This is another difference between the covariance matrices used here and the ones used in array processing.

More explicitly, when the two correlated noise signals are generated, their covariance matrix can be written in the form
\begin{equation}
	\expval{xx^T} = 
		\begin{bmatrix}
			R_{11} & R_{12}(0) \\
			R_{21}(0) & R_{22}
		\end{bmatrix}
\end{equation}
where each entry is a $2 \times 2$ block at time $t = 0$. (In the complex voltage representation, the entries are simply complex scalars.) We assume that the generated noise signals are stationary processes, so $R_{11}$ and $R_{22}$ do not vary in time. After signal 1 is transmitted, there is a time delay $\tau$ between the two signals when they are measured. Thus, at the time the signals are measured the covariance matrix becomes
\begin{equation} \label{eq:cov_blocks}
	\expval{xx^T} = 
	\begin{bmatrix}
		R_{11} & R_{12}(\tau) \\
		R_{21}(\tau) & R_{22}
	\end{bmatrix}\!\!.
\end{equation}
It is the $2 \times 2$ block matrix $R_{12}(\tau)$ that we use to perform matched filtering. This is because it encodes the correlation between the two signals.

In what follows, we will omit the time variable because, as described below, our experiment operates at a fixed range. In other words, $\tau$ is fixed for all of the results in this paper.

\section{Two-Mode Noise Radar as a Baseline for Comparison}
\label{sec:classical_setup}

As mentioned in the previous section, we have built a ``classical'' radar setup which works in a similar manner to our QTMS radar. Because it is not quite the same thing as a conventional noise radar, we have chosen to call it a \emph{two-mode noise radar} (TMN radar). Apart from the signal generation method, the two setups are identical. This allows the TMN radar to be a valid baseline for comparison with our QTMS radar.

\begin{figure}[h]
	\centerline{\includegraphics[width=\columnwidth]{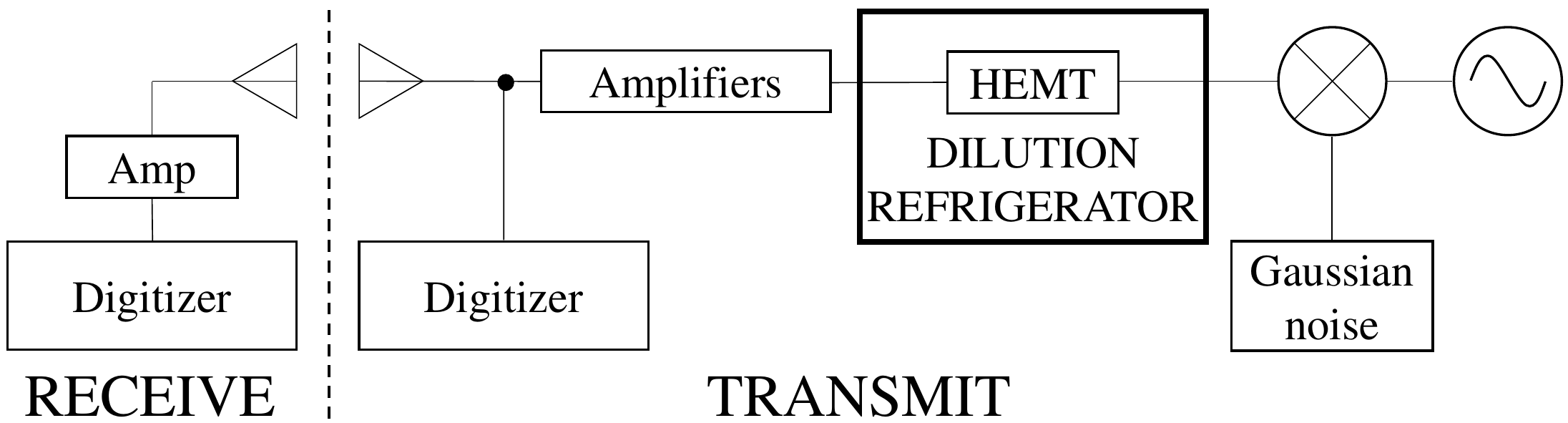}}
	\caption{Block diagram for the two-mode noise radar (TMN radar), a simple ``classical radar'' system which approximates the operation of our quantum radar prototype.}
	\label{fig:block_diagram_classical}
\end{figure}

A simple block diagram of the TMN radar setup is shown in Fig.\ \ref{fig:block_diagram_classical}. By comparing this to Figure \ref{fig:block_diagram_abstract}, we see that the basic structure has been retained but that the components are specified in more detail.

The signal generation step is the most important part. Here, we generate a carrier signal at 6.84105 GHz and mix it with Gaussian noise centered at 0.69655 GHz (band-limited, width 5 MHz). This produces two sidebands of correlated noise at 6.1445 GHz and 7.5376 GHz. After generation, the two sidebands are passed through a chain of amplifiers. Because we wish to treat the classical and quantum signals equally to the greatest possible extent, the amplifier chain is exactly the same in both cases. This means that the classical signal enters a dilution refrigerator, which cools its contents to cryogenic temperatures, to take advantage of the low-noise high-electron-mobility transistor (HEMT) amplifier within. More details on the cooling system will be found in Sec.\ \ref{sec:quantum_setup}; at this point it is only necessary to note that the signal is amplified.

The two noisy sidebands are fed into a splitter. One path ends in a digitizer which performs heterodyne measurements at 6.1445 GHz. This forms the measurement record to be used in the matched filter. The other path leads to an X-band horn antenna.

The receiver setup is extremely simple: it consists of another X-band antenna connected to an amplifier with a gain of 25 dB. This feeds into a digitizer performing measurements at 7.5376 GHz. The digitized measurement data is then correlated with the measurement record at 6.1445 GHz.

\begin{figure}[h]
	\centerline{\includegraphics[width=0.9\columnwidth]{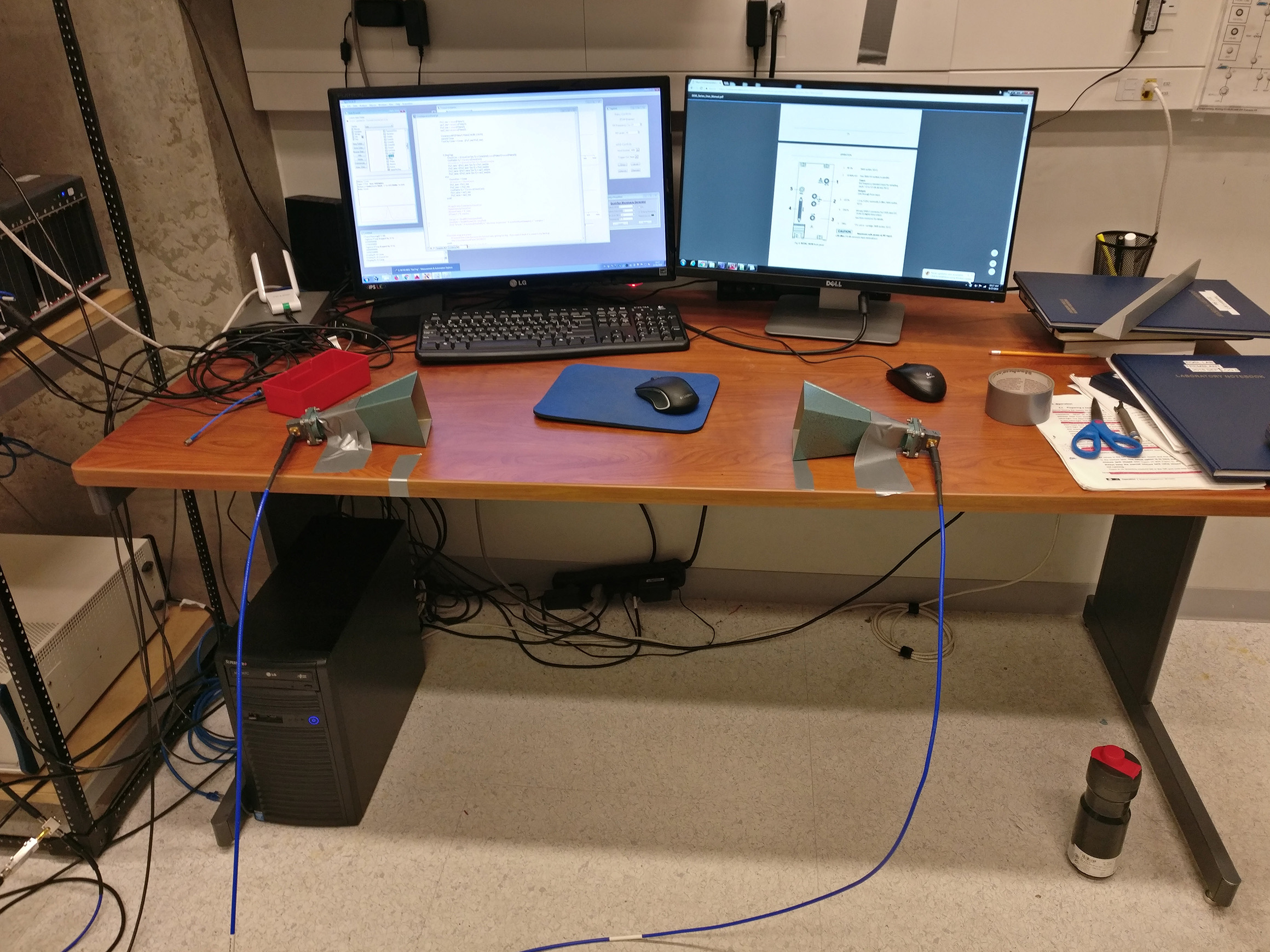}}
	\caption{The transmit and receive horns are mounted facing each other with a separation of 0.5 m.}
	\label{fig:horns_facing}
\end{figure}

The transmit and receive horns are mounted on a table, facing each other at a distance of 0.5 m (Fig. \ref{fig:horns_facing}). Our current arrangement is thus a one-way ranging setup rather than a true target detection apparatus. However, it does demonstrate the transmission of a signal through free space and the independence of the QTMS concept from the a priori need for range information. We have made no effort to isolate the horns or eliminate multipath.

The total power injected into the transmit horn is \textminus63 dBm. This includes a significant amount of noise from the amplifiers. The power of the signal originating from the Gaussian noise generator, after amplifier noise is discounted, is \textminus82 dBm. Based on this we could say, somewhat crudely, that the signal-to-noise ratio is \textminus19 dB. Given that our signal is Gaussian noise, we will not deal with signal-to-noise ratios in this paper because it is not always clear what exactly is signal, what exactly is noise, and how to quantify each. For example, we could argue that the true signal in this setup is not the transmitted 7.5376 GHz signal, but the covariance between this signal and the 6.1445 GHz signal. We could have chosen the definition of SNR which makes the quantum gain appear larger, but rather than that, we choose not to deal with SNR at all.

The return loss of the X-band horns was measured at 7.5 GHz to be 15 dB ($\approx$ 1.4 VSWR), which is acceptable for the purposes of this experiment.

\begin{table}
	\caption{Radar system parameters}
	\label{table:parameters}
	\centerline{
		\begin{tabular}{cc}  
			\toprule
			Parameter & Value \\
			\midrule
			Frequency 1 & 7.5376 GHz \\
			Frequency 2 & 6.1445 GHz \\
			Bandwidth & 1 MHz \\
			Power at Tx antenna input & \textminus82 dBm \\
			Return loss (at 7.5 GHz) & 15 dB (1.4 VSWR) \\
			Antenna gain (at 7.5 GHz) & 15 dB \\
			Vertical beam width & $\approx 27^\circ$ \\
			Horizontal beam width & $\approx 37^\circ$ \\
			Sampling frequency & 1 MHz \\
			\bottomrule
		\end{tabular}
	}
\end{table}

A summary of the basic parameters of the TMN radar is given in Table \ref{table:parameters}. The same parameters also hold for our QTMS radar prototype.

At present, both frequencies are transmitted, though the 6.1445 GHz signal is attenuated because it is out of the passband for the X-band antenna. In future work, we intend to install C-band antennas to better match our operating frequencies and improve the return loss. We also intend to filter out the 6.1445 GHz component, possibly by installing a filter at the splitter so that only the 7.5376 GHz signal is transmitted.

Note that, though not explicitly stated, the 6.1445 GHz and 7.5376 GHz signals have to be demodulated to baseband or IF before being correlated for the generation of statistics (as described in Sec.\ \ref{sec:abstract_setup}). That being said, we could have simply correlated the demodulated 7.5376 GHz with the original Gaussian signal used to generate the two sidebands. However, the point is to retain an architecture which is parallel to the QTMS radar architecture to be described in Sec.\ \ref{sec:quantum_setup}. Hence the TMN radar described here is not perfectly classical, but is tailored to ensure apples are being compared with apples when the classical signal statistics are compared to the quantum signal statistics.

\subsection{Covariance Matrix for the TMN Radar}

In this subsection, we perform a rudimentary analysis in order to derive the form of the covariance matrix generated from the signals in the TMN radar described above.

Recall that we generate a carrier signal and mix it with Gaussian noise, which we assume is a bandpass process with bandwidth $B$. If the Gaussian noise has a center frequency of $\omega$ such that $\omega > B/2$, we may express it in terms of in-phase and quadrature components as
\begin{equation} \label{eq:complex_signal}
	X(t) = [X_I(t) + j X_Q(t)] [ \cos(\omega t) + j \sin(\omega t) ]
\end{equation}
where $X_I(t)$ and $X_Q(t)$ are uncorrelated, zero-mean Gaussian processes, each with variance $\sigma^2$. In the following calculation we are only interested in the physical signal being transmitted, which is analogous to the real part of \eqref{eq:complex_signal}:
\begin{equation} \label{eq:real_signal}
	\operatorname{Re}[X(t)] = X_I(t) \cos(\omega t) - X_Q(t) \sin(\omega t).
\end{equation}
In any case, the complex signal can always be recovered from its real part \cite{therrien1992discrete}. As for the carrier signal, we may write it as
\begin{equation}
	c(t) = 2 \cos(\omega_0 t)
\end{equation}
where $\omega_0$ is the angular frequency of the signal. (We have chosen an amplitude of 2 to eliminate certain fractions below; this does not lead to a loss of generality.)

Assuming that the mixer is ideal and that $\omega_0 > \omega$, the resultant signal is the product of the carrier and the noise signal:
\begin{align}\label{eq:mixed_signal}
\begin{split}
	\operatorname{Re}[c(t) X(t)] &= 2 X_I(t) \cos(\omega_0 t) \cos(\omega t) \\
	 	&\phantom{=\ } - 2 X_Q(t) \cos(\omega_0 t) \sin(\omega t) \\
		&= X_I(t) \cos(\omega_1 t) - X_Q(t) \sin(\omega_1 t) \\
		&\phantom{=\ } + X_I(t) \cos(\omega_2 t) - [-X_Q(t)] \sin(\omega_2 t)
\end{split}
\end{align}
where we define $\omega_1 = \omega_0 + \omega$ and $\omega_2 = \omega_0 - \omega$. We can see that, as stated above, there are two sidebands centered on the angular frequencies $\omega_1$ and $\omega_2$. In the perfect case, these would be the signals received at the digitizers (with a possible corresponding reduction in amplitude based on whether it was transmitted through free space or directly measured after generation).

In the event that a phase shift occurs between the two signals, we incorporate this by adding a phase to the $\omega_1$ component of the signal in \eqref{eq:mixed_signal}:
\begin{equation}\label{eq:phase_shift}
\begin{split}
	&X_I(t) \cos(\omega_1 t + \phi) - X_Q(t) \sin(\omega_1 t + \phi) \\
	&\qquad= \bigl[ X_I(t) \cos\phi - X_Q(t) \sin\phi \bigr] \cos(\omega_1 t) \\
	&\qquad\phantom{=\ } - \bigl[ X_I(t) \sin\phi + X_Q(t) \cos\phi \bigr] \sin(\omega_1 t).
\end{split}
\end{equation}

In a realistic system there is always noise, whether from within the system or the environment. To model this noise, which we assume to be additive white Gaussian noise (AWGN), we introduce zero-mean Gaussian processes $n_{1I}(t)$, $n_{1Q}(t)$, $n_{2I}(t)$, and $n_{2Q}(t)$. The first two have a variance of $\sigma_{n1}^2$; the latter two a variance of $\sigma_{n2}^2$. Then, following \eqref{eq:real_signal}, the real parts of the signals received at the digitizers, $\operatorname{Re}[X_1(t)]$ and $\operatorname{Re}[X_2(t)]$, are
\begin{subequations} \label{eq:classical_signals}
\begin{align}
\begin{split}
	&\operatorname{Re}[X_1(t)] = \\
		&\qquad A_1 \bigl[ X_I(t) \cos\phi - X_Q(t) \sin\phi + n_{1I}(t) \bigr] \cos(\omega_1 t) \\
		&\qquad - A_1 \bigl[ X_I(t) \sin\phi + X_Q(t) \cos\phi + n_{1Q}(t) \bigr] \sin(\omega_1 t)
\end{split} \\
\begin{split}
	&\operatorname{Re}[X_2(t)] = A_2 \bigl[ X_I(t) + n_{2I}(t) \bigr] \cos(\omega_2 t) \\
		&\phantom{\operatorname{Re}[X_2(t)] =\ } - A_2 \bigl[ -X_Q(t) + n_{2Q}(t) \bigr] \sin(\omega_2 t).
\end{split}
\end{align}
\end{subequations}
These were obtained by separating out the $\cos(\omega_1 t)$, $\sin(\omega_1 t)$, $\cos(\omega_2 t)$, and $\sin(\omega_2 t)$ components of \eqref{eq:mixed_signal} and \eqref{eq:phase_shift}, then adding Gaussian processes to each component to represent noise. The factors $A_1$ and $A_2$ take into account any amplifier gains together with all losses due to system components, distance between the antennas, etc.

The next step is to demodulate to baseband; note that this has to be done using two separate digitizers measuring at $\omega_1$ and $\omega_2$. Once this is done, a comparison with \eqref{eq:real_signal} shows that the equations in \eqref{eq:classical_signals} can be seen to represent the measured quadrature voltages as follows:
\begin{subequations}\label{eq:classical_IQ}
	\begin{align}
	I_1(t) &= A_1 \bigl[ X_I(t) \cos\phi - X_Q(t) \sin\phi + n_{1I}(t) \bigr] \\
	Q_1(t) &= A_1 \bigl[ X_I(t) \sin\phi + X_Q(t) \cos\phi + n_{1Q}(t) \bigr] \\
	I_2(t) &= A_2 \bigl[ X_I(t) + n_{2I}(t) \bigr] \\
	Q_2(t) &= A_2 \bigl[ -X_Q(t) + n_{2Q}(t) \bigr].
	\end{align}
\end{subequations}

To complete the calculation, recall that $\expval{X_I(t) X_Q(t)} = 0$, that the introduced noise is independent of the generated signal, and that all the processes involved have zero mean. Once this is done, we find that the covariance matrix for our TMN radar is
\begin{align} \label{eq:classical_cov}
	\expval{xx^T} = 
	\begin{bmatrix}
		\sigma_1^2 & 0 & \rho s \cos\phi & \rho s \sin\phi \\
		0 & \sigma_1^2 & \rho s \sin\phi & -\rho s \cos\phi \\
		\rho s \cos\phi & \rho s \sin\phi & \sigma_2^2 & 0 \\
		\rho s \sin\phi & -\rho s \cos\phi & 0 & \sigma_2^2
	\end{bmatrix}
\end{align}
where $x = [I_1, Q_1, I_2, Q_2]^T$ and
\begin{subequations}\label{eq:sigmarho}
\begin{align}
	\sigma_1^2 &= A_1^2 (\sigma^2 + \sigma_{n1}^2) \\
	\sigma_2^2 &= A_2^2 (\sigma^2 + \sigma_{n2}^2) \\
	\rho &= \left[ \left( 1 + \frac{\sigma_{n1}^2}{\sigma^2} \right) \!\! \left( 1 + \frac{\sigma_{n2}^2}{\sigma^2} \right) \right]^{\! -\frac{1}{2}}.
\end{align}
\end{subequations}
We write the matrix in this form so we can explicitly give an expression for $\rho$, which is the Pearson correlation coefficient. Recall that $\sigma^2$ was the variance of the in-phase and quadrature components of the original Gaussian noise signal, defined at the beginning of this subsection. We define $s = \sigma_1\sigma_2$ purely as a space-saving measure. This matrix may be compared to the general block-matrix form given in \eqref{eq:cov_blocks}.

Note the negative sign in the $\expval{Q_1 Q_2}$ entry of the matrix, showing that the two quadratures are anticorrelated. This is a consequence of the signal generation process and distinguishes our TMN radar from conventional noise radars, in which the transmitted and recorded signals are identical and the two quadratures would be positively correlated. Compare Eq.\ 13 of \cite{dawood2001roc}, in which $\expval{Q_1 Q_2}$ is positive (though they use a different notation).

\section{Quantum Noise and Entanglement}

The success of any radar based on the abstract scheme described in Sec.\ \ref{sec:abstract_setup} depends largely on the \emph{correlation} between the two generated noise signals. Higher correlation means better robustness against extraneous noise. The goal of our QTMS radar is to increase this correlation relative to the ``classical'' TMN radar described in the previous section. In order to do this, we take advantage of a quantum phenomenon called \emph{entanglement}. Before we explain this, however, we must first explain the concept of \emph{quantum noise}.

According to classical electromagnetism, it is possible for two independently generated light waves to be perfectly correlated if they share the exact same waveform. This is impossible to realize in practice, of course: there will always be noise somewhere in the system. There is no obstacle in the classical theory, however, which prevents this noise from being eliminated entirely. Astonishingly, when we pass to the more fundamental description given by quantum electrodynamics, we find that \emph{even in theory, the noise cannot be reduced to zero}. This noise stems from the quantum nature of light. It turns out that in-phase ($I$) and quadrature ($Q$) voltages of any signal cannot be simultaneously measured with infinite precision. In the appropriate units---the specifics are unimportant---they satisfy the inequality
\begin{equation}
	\sigma_I^2 \sigma_Q^2 \geq \frac{1}{4}
\end{equation}
where $\sigma_I^2$ and $\sigma_Q^2$ are the variances in $I$ and $Q$ respectively \cite{adesso2014cv}. This is an analog of the famous Heisenberg uncertainty principle, but applied to quadrature components rather than position and momentum. (Expressions like this are part of the reason why real-valued voltages, rather than complex ones of the form $I + jQ$, are more natural in quantum physics.)

\begin{figure}[t]
	\centering
	\subfloat[]{\includegraphics[width=.9\columnwidth]{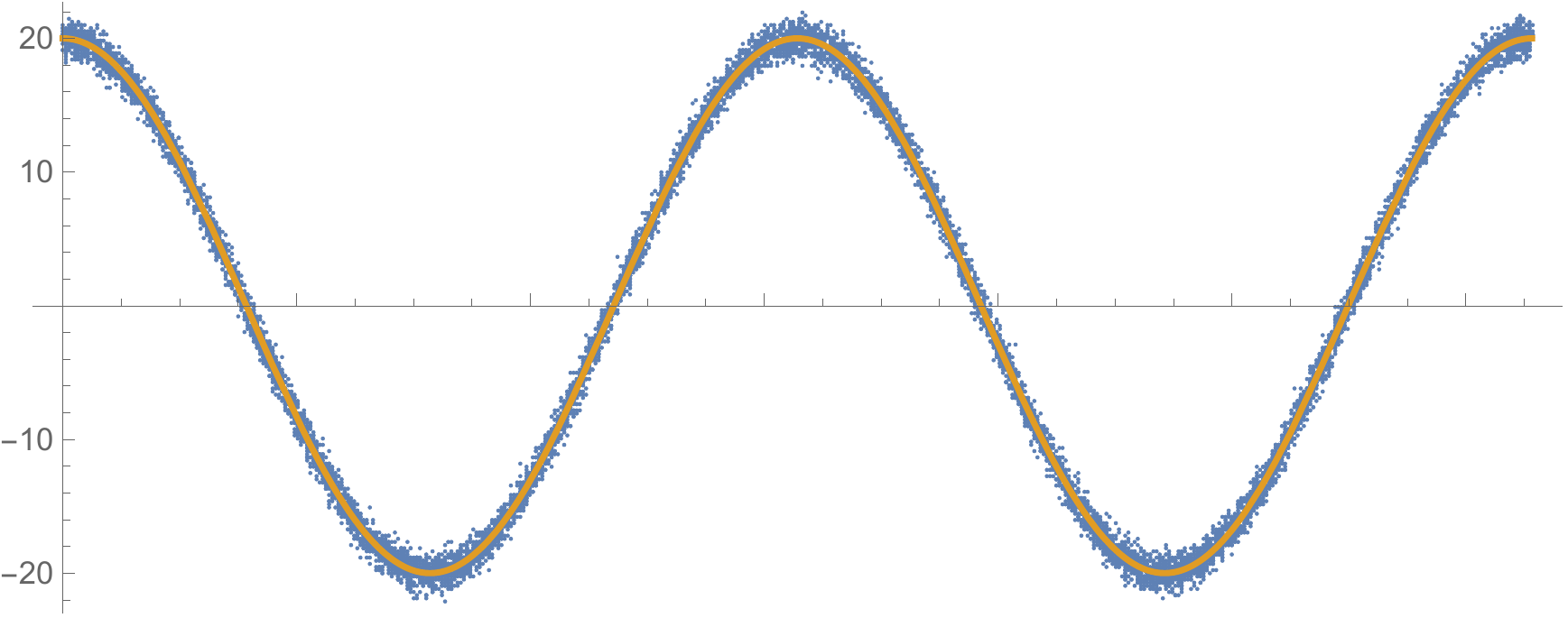}
		\label{subfig:coherent_state_2}}
	\hfil
	\subfloat[]{\includegraphics[width=.9\columnwidth]{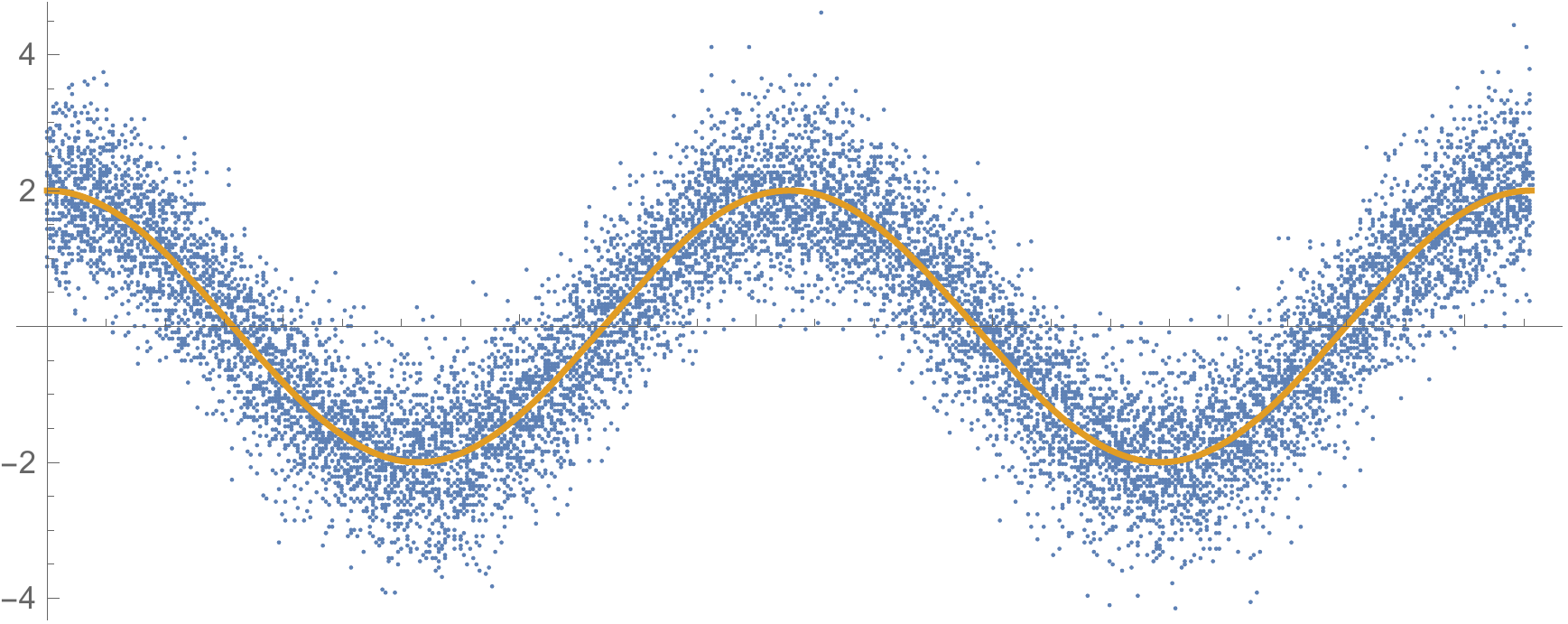}
		\label{subfig:coherent_state_20}}
	\caption{Solid curves: an ideal sinusoid waveform in one of the quadrature voltages as a function of time (or, equivalently, of phase). Dots: inherent quantum noise in the quadrature. Plot (b) shows a sinusoid with an amplitude 10 times as small as that in plot (a), from which it can be seen that quantum noise becomes significant at low signal powers. Units are arbitrary.}
	\label{fig:coherent_state}
\end{figure}

An illustration of this can be seen in Fig.\ \ref{fig:coherent_state}, which shows two sinusoid waveforms and their corresponding ``coherent states'', the closest quantum mechanical analogs. (Lasers in the optical regime, for example, are described by sinusoids in classical electromagnetism and by coherent states in quantum optics.) Suppose we have a waveform whose in-phase quadrature voltage is described by $I_\text{classical}(t) = A \cos(\omega t)$, where $A$ is the amplitude and $\omega$ is the angular frequency. There is nothing in classical physics to suggest that this exact waveform cannot be generated by some physical system. In quantum mechanics, however, this cannot be achieved by any physical system, no matter how ideal. In the theoretically ideal case, measurements of $I$ voltages are actually \emph{random}, with a Gaussian distribution centered around the classical value $I_\text{classical}(t)$. Thus $I(t)$, the time series of measured $I$ voltages, is a random variable:
\begin{equation}\label{eq:coherent_state_noise}
	I(t) \sim I_\text{classical}(t) + \mathcal{N}(0,\sigma_I^2)
\end{equation}
The noise $\mathcal{N}(0,\sigma_I^2)$ is seen to have a mean of 0. The variance $\sigma_I^2$ is a function of $\omega$, but is independent of the amplitude $A$. This is an important observation: it means that as the amplitude decreases, the importance of quantum noise increases. This can also be seen by comparing the two plots in Fig.\ \ref{fig:coherent_state}. Although we have chosen this example merely for illustrative purposes, this gives us reason to believe that quantum radars which compensate for the quantum noise could be useful in the low-power regime. In fact, our QTMS radar does not use coherent states but the same intuition applies. (This does not mean quantum techniques will \emph{always} lead to an improvement over classical radars: different classes of quantum signals yield different results.)

We reiterate that the noisiness described by \eqref{eq:coherent_state_noise} and seen in Fig.\ \ref{fig:coherent_state} is \emph{unavoidable}; it is the theoretical minimum of noise that any light wave must contain according to quantum theory. The scale of the quantum noise is $hf/2$ per Hertz of bandwidth (in other words, half a photon per Hertz), where $h$ is the Planck constant and $f$ is the center frequency of the signal under consideration.

This means that, in a radar like the TMN radar described in Sec.\ \ref{sec:classical_setup}, the transmitted and recorded signals can never be perfectly correlated. They are subject to the laws of quantum mechanics and therefore contain quantum noise. In particular, each sideband produced by the mixer (see Fig.\ \ref{fig:block_diagram_classical}) contains, in addition to the classical Gaussian noise, an added quantum noise component. Only the classical noise is correlated between the sidebands; the added quantum noise is independent.

Thus, quantum theory puts bounds on \eqref{eq:classical_cov} whose existence we would never have suspected if we restricted ourselves to classical electromagnetism. It tells us that the noise terms $n_{1I}(t)$, $n_{Q1}(t)$, $n_{2I}(t)$, and $n_{2Q}(t)$ can never go to zero because they each contain a contribution from quantum noise. This reduces the correlation $\rho$ between the two signals, particularly at low power levels. For more details, see \cite{chang2018multimode,chang2018quantum}.

This is where \emph{entanglement} comes into play. It is possible to create pairs of signals in which the \emph{quantum noise} is correlated. If two signals exhibit higher correlations than that achievable when the quantum noise is uncorrelated, the two signals are said to be \emph{entangled}. (Note that this is only one type of entanglement, which we call \emph{continuous variable entanglement} because the correlations are between continuous variables like voltage, as opposed to discrete variables like polarization.) Thus entanglement boosts the level of correlation between the transmitted and recorded signals, leading to a quantum enhancement. In terms of \eqref{eq:classical_cov}, we can roughly say that our QTMS radar produces signals with a higher correlation coefficient $\rho$ when compared to our TMN radar.

Note, however, that when we characterize entanglement as correlations superior to those achievable when ignoring quantum noise, we are not giving a \emph{definition} of entanglement. This is only an operational description which we choose to use because it is easy to understand from a radar engineering viewpoint. A rigorous treatment of entanglement is beyond the scope of this paper. On the other hand, the above explanation contains everything needed to understand our work. Therefore, even though the concept of entanglement may seem shrouded in mystery, its application to QTMS radar is relatively simple. We need not consider abstruse principles like nonlocality. It suffices to understand that quantum noise, inherent in all $I$ and $Q$ measurements, is uncorrelated in classically-generated signals but correlated in entangled signals.

\subsection{The Entangled Signal: Two-mode Squeezed Vacuum}
\label{subsec:TMSV}

The entangled signal we use in our prototype radar is called \emph{two-mode squeezed vacuum} (TMSV). This consists of two beams at different frequencies---that is, two frequency \emph{modes}---which are entangled with each other such that they satisfy the following property: when we generate a series of TMSV pulses and measure the quadrature voltages $I_1$, $Q_1$, $I_2$, and $Q_2$, the measurement results follow a multivariate Gaussian distribution with mean and covariance matrix as given below. (As is well known, a multivariate Gaussian distribution is completely specified by its mean and covariance matrix.) In the quantum optics literature, TMSV signals are classed among the so-called ``Gaussian states'' because of this behavior.

Note that TMSV signals \emph{do not have an associated waveform}. Indeed, the notion of ``waveform'' is not used in quantum mechanics because waveforms, being a classical concept, cannot be used to describe quantum mechanical behavior. Certain quantum signals, like the ``coherent states'' described above, have \emph{average} waveforms, but such waveforms are not an \emph{exact} quantum mechanical description. There is no such approximation for TMSV signals because they have mean zero. They can be thought of like a stochastic process, which is why we must describe them in the probabilistic terms used in the previous paragraph.

For ideal TMSV signals, all the measured quadrature voltages have zero mean:
\begin{equation} \label{eq:TMSV_zeromean}
	 \expval{I_1,Q_1,I_2,Q_2}^T = [0,0,0,0]^T.
\end{equation}

The covariance matrix for ideal TMSV signals is, up to a constant of proportionality which depends on the units and conventions chosen,
\begin{equation} \label{eq:TMSV_cov}
	\expval{xx^T} = \begin{bmatrix}
		C & 0 & S \cos\phi & S \sin\phi \\
		0 & C & S \sin\phi & -S \cos\phi \\
		S \cos\phi & S \sin\phi & C & 0 \\
		S \sin\phi & -S \cos\phi & 0 & C
	\end{bmatrix}
\end{equation}
where we have defined
\begin{subequations}
	\begin{align}
	C &= \cosh 2r \\
	S &= \sinh 2r
	\end{align}
\end{subequations}
to save space. The real numbers $r$ and $\phi$ are the magnitude and phase, respectively, of the complex parameter $r e^{j \phi}$. This is known as the \emph{squeezing parameter}.

We will not present the derivation of \eqref{eq:TMSV_cov} in this paper because it requires significant background knowledge in quantum physics. The interested reader may consult \cite{braunstein2005qi}, which however gives this matrix only for the case $\phi = 0$. In the context of entanglement, this phase is unimportant and we can always choose $\phi = 0$; we have restored the phase to show the connection with \eqref{eq:classical_cov}.

Since the measured voltages of TMSV signals are random, as noted above, we can say that the two beams consist purely of quantum noise. There is no need for a separate noise generator, Gaussian or otherwise. What is more, it can be shown that each individual beam is indistinguishable from thermal noise. Only by correlating the two beams can we see the correlations. In a radar context, this means that there is no distinctive signature that an adversary can use to distinguish a QTMS radar beam from background thermal noise.

In the term ``two-mode squeezed vacuum'', the word ``squeezed'' means that the quantum noise in certain linear combinations of quadratures are reduced---squeezed---at the price of increasing the noise in other combinations. For example, when $\phi = 0$ we find that $(I_1 - I_2)/\sqrt{2}$ and $(Q_1 + Q_2)/\sqrt{2}$ are squeezed: they each have a variance of $e^{-2r}$. To compensate for this, $(I_1 + I_2)/\sqrt{2}$ and $(Q_1 - Q_2)/\sqrt{2}$ have increased noise: they each have variance $e^{2r}$ \cite{braunstein2005qi,drummond2004squeezing}. This can be verified for any $\phi$ by diagonalizing the covariance matrix \eqref{eq:TMSV_cov}, but when $\phi = 0$ we can perform a direct calculation. For example,
\begin{align*}
	\variance \! \left[ \frac{I_1 - I_2}{\sqrt{2}} \right] &= \frac{1}{2} \big( \! \expval{(I_1 - I_2)^2} - \expval{(I_1 - I_2)}^2 \big) \\
		&= \frac{1}{2} \big( \! \expval{I_1^2} - 2 \expval{I_1 I_2} + \expval{I_2^2} \\
		&\phantom{= } - \expval{I_1}^2 + 2 \expval{I_1} \expval{I_2} - \expval{I_2}^2 \big) \\
		&= \frac{1}{2} \big( \! \variance[I_1] + \variance[I_2] - 2 \expval{I_1 I_2} \big) \\
		&= e^{-2r}.
\end{align*}
In this calculation we used \eqref{eq:TMSV_zeromean} and the relevant entries of \eqref{eq:TMSV_cov}. This exponentially decreasing variance is why $r e^{j \phi}$ is called the squeezing parameter.

A comparison of \eqref{eq:classical_cov} and \eqref{eq:TMSV_cov} shows that the two covariance matrices have essentially the same structure. Indeed, by setting $\sigma_1^2 = \sigma_2^2 = \cosh 2r$ and $\rho = \tanh 2r$ in \eqref{eq:classical_cov}, we obtain \eqref{eq:TMSV_cov}. This shows that the TMN radar described in Sec.\ \ref{sec:classical_setup} is indeed a reasonable approximation to the operation of our QTMS radar. Note, however, that this mathematical correspondence does not mean that the two are physically the same: only an entangled signal generator can generate signals with covariances described by \eqref{eq:TMSV_cov}.

The entanglement of a given TMSV signal generator can be verified from the estimated covariance matrix. For example, one condition is that the off-diagonal blocks $R_{12}$ and $R_{21}$ in \eqref{eq:cov_blocks} have negative determinant. (This is also fulfilled by TMN radar, which is one reason why we chose it as a standard of comparison.) A detailed discussion is beyond the scope of this paper; it is sufficient to state that we have performed tests to check that our generated signal was entangled at the source (that is, before amplification or transmission). We describe them in \cite{chang2018multimode}. However, we should point out that very precise measurements are required in order to verify entanglement. This requires extremely careful calibration of the experimental system so that the powers and correlations emanating from the entanglement source are known to the level of single photons. The amount of noise added by the system is also important, because the added noise may degrade the correlations to the point that, somewhere along the line, the signals cease to be entangled (though they are still correlated). Therefore it is necessary to take this noise into account when checking whether a signal generator is generating entangled signals. To date, this level of precision is not necessary, nor is it even approached, in practical radar systems.

\section{Experimental Setup}
\label{sec:quantum_setup}

\begin{figure*}[t]
	\centerline{\includegraphics[width=.75\textwidth]{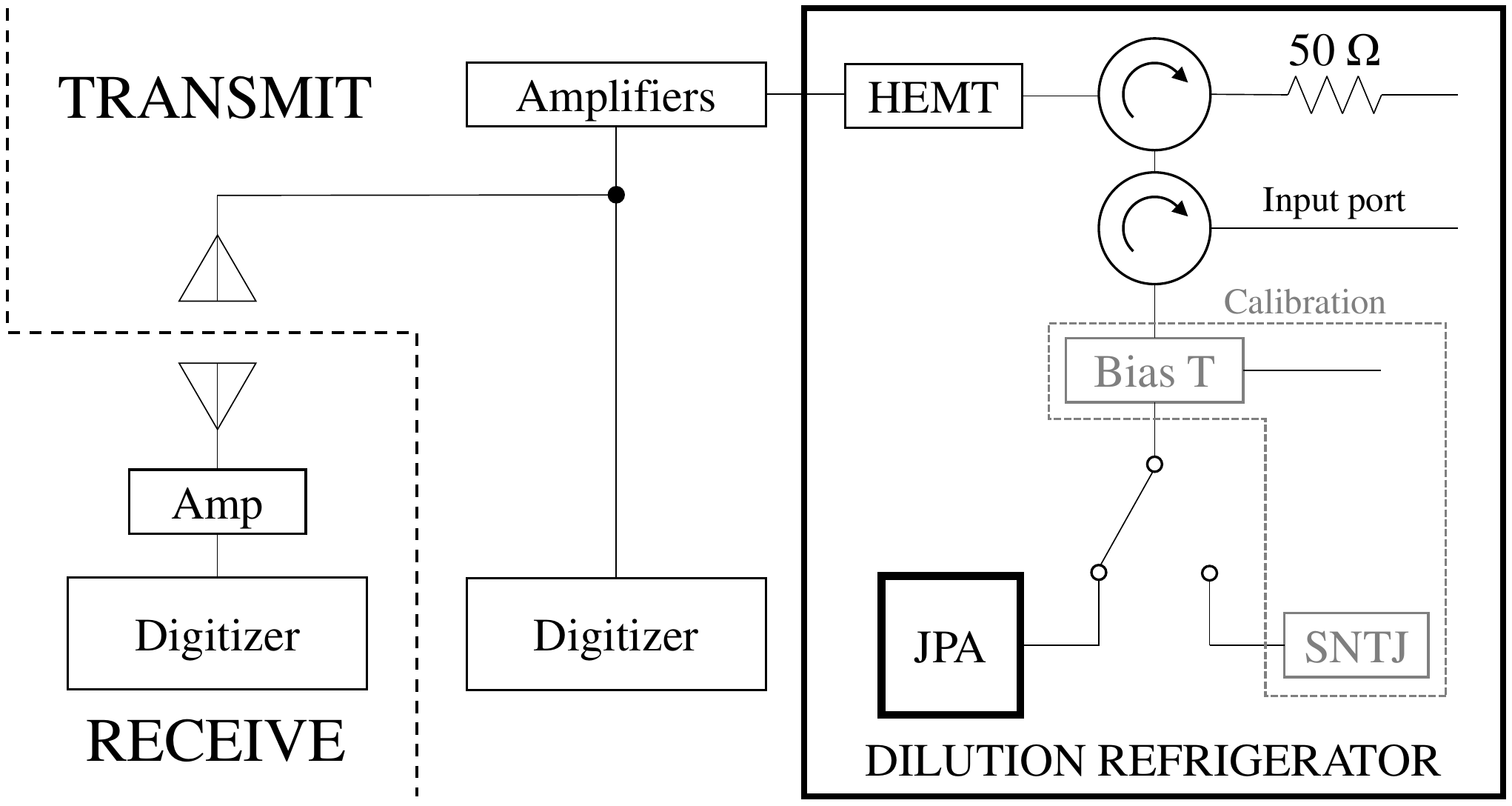}}
	\caption{Simplified block diagram of our quantum radar setup. The JPA generates the entangled signal, which undergoes amplification and is transmitted through a horn antenna.}
	\label{fig:block_diagram_quantum}
\end{figure*}

A simplified block diagram of our QTMS radar setup is shown in Fig.\ \ref{fig:block_diagram_quantum}. Once the signal emerges from the dilution refrigerator, the setup is exactly the same as in the TMN radar described in Sec.\ \ref{sec:classical_setup}. This can be seen by comparing Figures \ref{fig:block_diagram_classical} and \ref{fig:block_diagram_quantum}. Therefore, the values in Table \ref{table:parameters} apply to our QTMS radar prototype as well. 

The only difference between our QTMS radar and the classical TMN radar is in the signal generation step. In the classical case the desired sidebands were created by mixing a carrier signal with band-limited Gaussian noise. TMSV signals are not so easily generated, and we thus have a more complicated system.

Be that as it may, there are really only two aspects of our setup which would be unusual to a radar engineer: the Josephson parametric amplifier (JPA) and the dilution refrigerator. The JPA is the source of the TMSV entangled signal. It is the most important part of our system and is the only component that is not commercially available. We will examine it in more detail in the next section.

\begin{figure}[t]
	\centerline{\includegraphics[width=0.8\columnwidth]{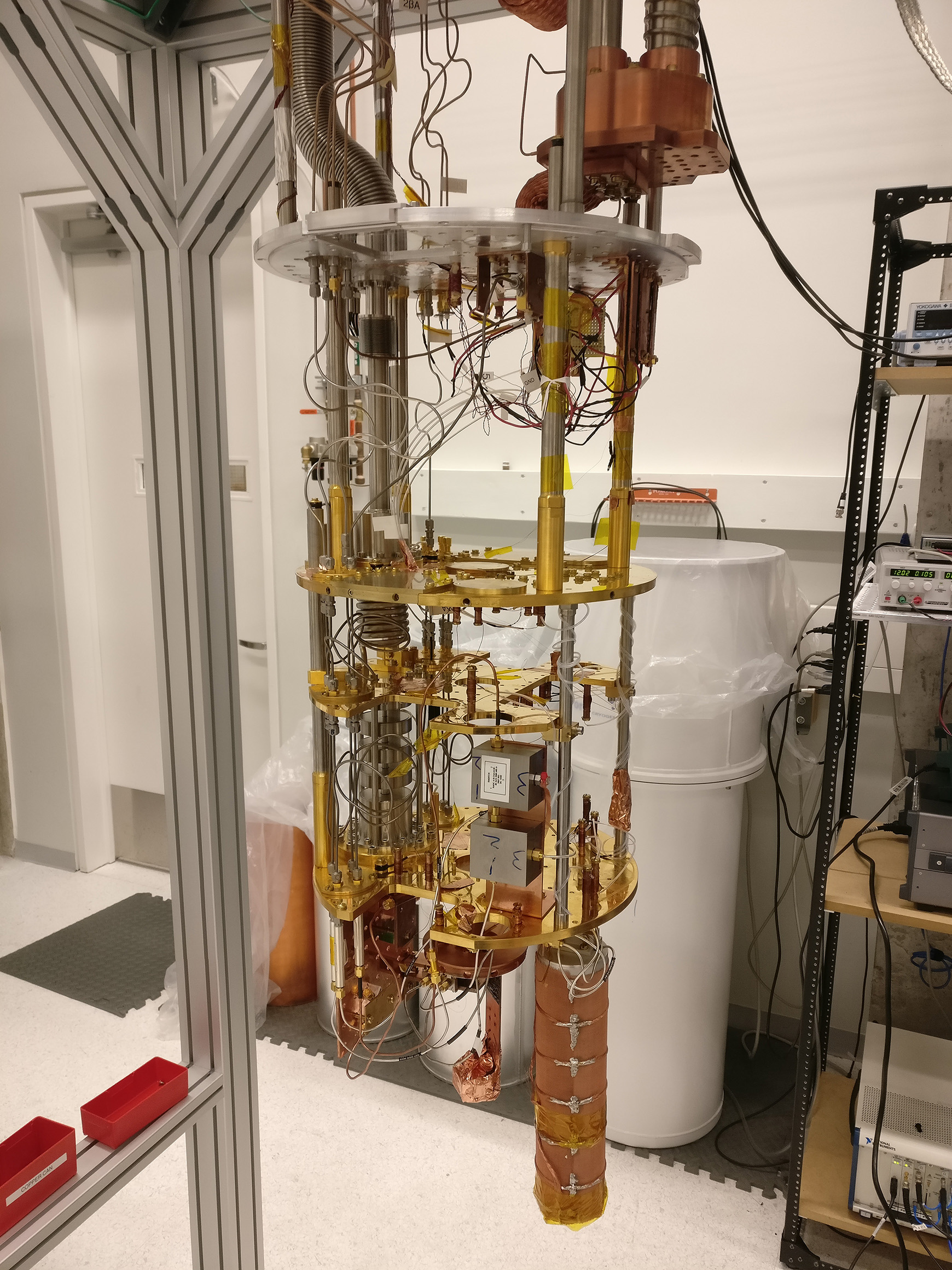}}
	\caption{Interior of the dilution refrigerator. From top to bottom, each round plate is colder than the last. The JPA is inside the can at the bottom.}
	\label{fig:fridge_inside_overall}
\end{figure}

\begin{figure}[t]
	\centerline{\includegraphics[width=0.8\columnwidth]{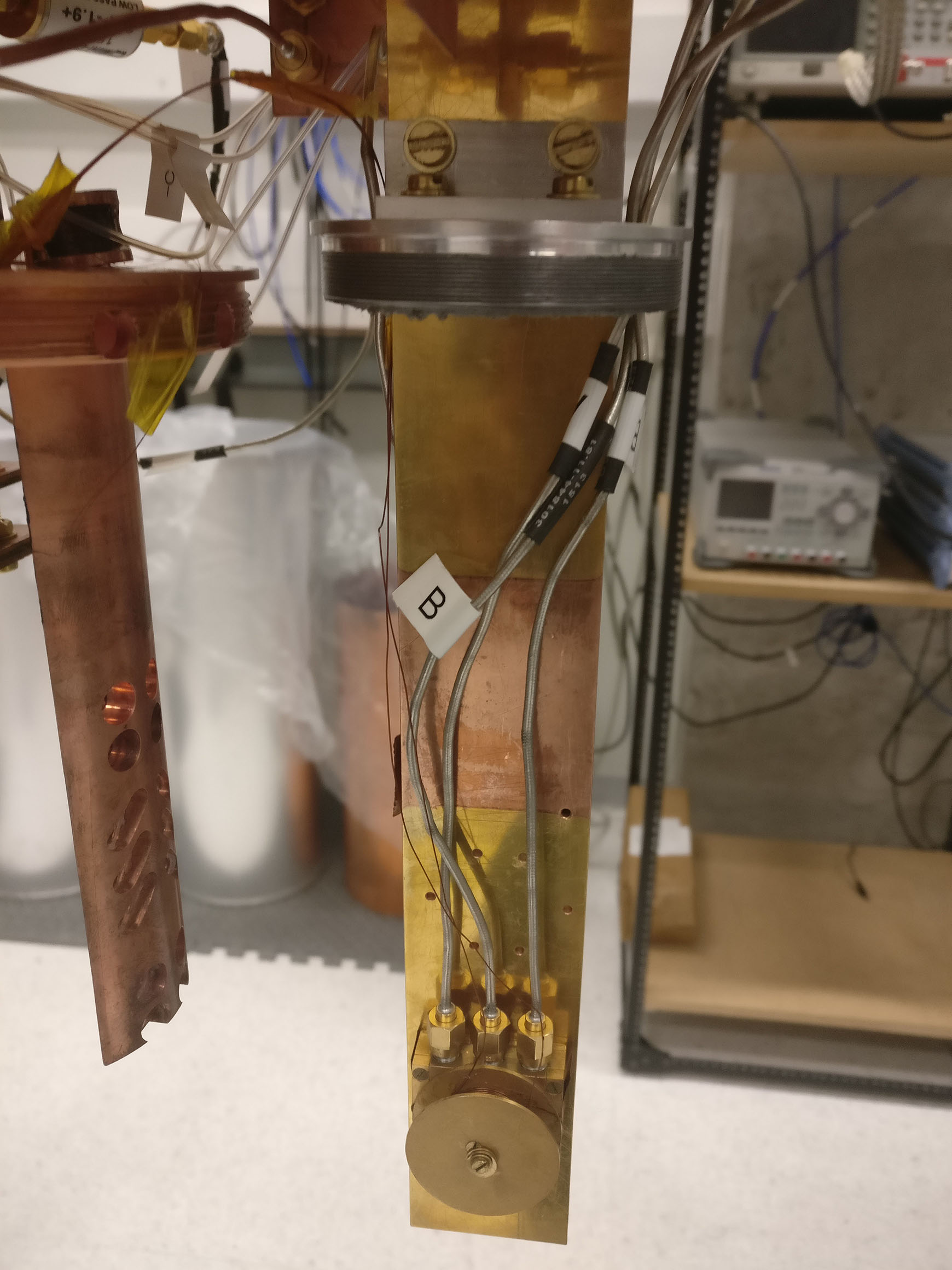}}
	\caption{Interior of the can containing the JPA (see Fig.\ \ref{fig:fridge_inside_overall}). The JPA (inside the square box near the bottom, behind the circular coil) is mounted in the coldest stage of the dilution refrigerator. The coil just outside the JPA produces a static magnetic field. This picture was reused from \cite{luong2018mqr}.}
	\label{fig:fridge_JPA}
\end{figure}

The dilution refrigerator (Fig. \ref{fig:fridge_inside_overall}), manufactured by BlueFors Cryogenics, uses liquid helium to cool its contents to cryogenic temperatures. It contains several stages, each colder than the last. The JPA itself is located in the coldest stage of the cryostat (Fig. \ref{fig:fridge_JPA}), which is cooled to approximately 7 mK. The device needs to be very cold because the JPA is a superconducting device, but the reason we go as low as 7 mK is to produce an electromagnetic vacuum at frequencies above 4 GHz. This is needed because the JPA is extremely sensitive to thermal noise.

The JPA is connected through a microwave switch to a bias tee, two circulators, and a chain of amplifiers beginning with a high-electron-mobility transistor (HEMT). Apart from the HEMT itself, the amplifier chain is at room temperature. (We use exactly the same chain of amplifiers for the TMN radar; see Sec.\ \ref{sec:classical_setup}.) After the signal is amplified, it is split between a horn antenna and a digitizer in the same way as in the TMN radar setup (Sec.\ \ref{sec:classical_setup}).

The outgoing signal undergoes amplification so that it can be more easily detected, just as in the classical case. The HEMT is used as a low-noise amplifier because, as is well known, the added noise in a chain of amplifiers is dominated by the noise from the first amplifier \cite{kingsley2016radar}. The amount of entanglement is not increased by this amplification process; on the contrary, it is \emph{destroyed} by the added noise. However, a quantum enhancement (in the form of higher correlations) persists, as shown in our experimental results.

The two circulators are used as isolators to prevent any extraneous signal from reaching the JPA. They may also be used to appropriately direct an input signal if this were desired, though this is unnecessary for our quantum radar prototype.

The microwave switch can be toggled between the JPA and a device called a shot-noise tunnel junction (SNTJ). This is used together with the bias tee as part of a calibration process to confirm that the JPA signal is entangled. This calibration is performed approximately once a day. More details of this calibration and testing process are given in \cite{chang2018multimode}. Using this process, we have confirmed that at the time we performed our quantum radar experiments, the signal emanating from the JPA was entangled. We are confident, therefore, that our prototype is indeed a quantum radar in some sense. The SNTJ and bias tee play no other role in our system.

\section{Entanglement Generation Using a Josephson Parametric Amplifier}
\label{sec:JPA}

\begin{figure}[h]
	\centerline{\includegraphics[width=.9\columnwidth]{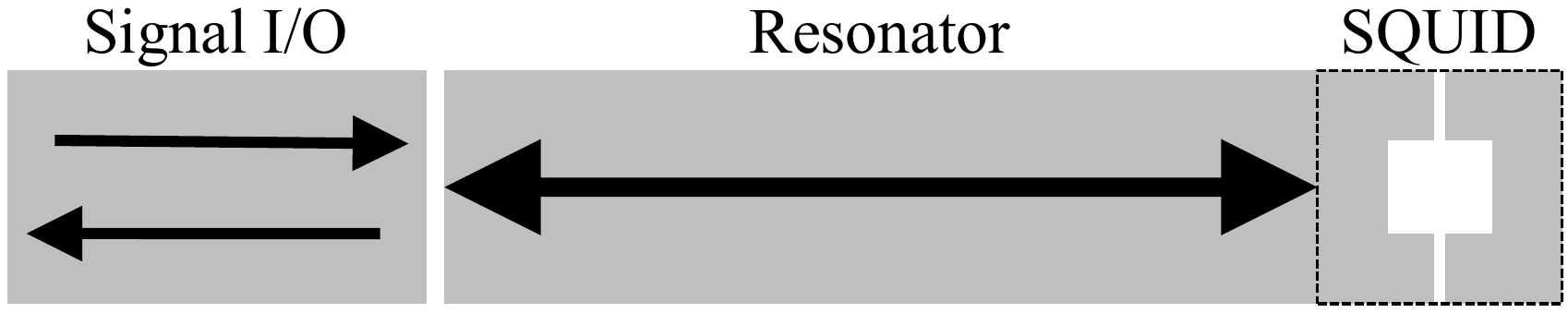}}
	\caption{Simplified diagram of a JPA. This figure was reused from \cite{luong2018mqr}.}
	\label{fig:JPA_schematic}
\end{figure}

At the heart of our QTMS radar is a \emph{Josephson parametric amplifier} (JPA). This is the device that generates the entangled TMSV signal. A JPA is essentially a microwave resonator with a variable resonance frequency. The type of JPA we used consists of a resonant cavity with a superconducting quantum interference device (SQUID) at one end (Fig. \ref{fig:JPA_schematic}). The resonance frequencies of the cavity can be modified by applying a magnetic field to the SQUID. If the magnetic field is modulated at a frequency corresponding to the sum of two frequencies, the JPA can perform \emph{parametric amplification} of signals incident on the cavity. Parametric amplification can be explained classically, but when the input signal is the quantum vacuum it produce the entangled TMSV signal. The ``vacuum'' part of the name ``two-mode squeezed vacuum'' comes from this fact.

It may be helpful to make an analogy with an optical cavity. The mirrors in an optical cavity correspond to impedance mismatches at either end of the JPA. Moving one of the mirrors changes the resonant frequencies of the optical cavity; exposing the SQUID to a magnetic field does the same in the JPA. (In effect, a magnetic field changes the impedance of the SQUID.) An oscillating magnetic field corresponds to a vibrating mirror.

In our QTMS radar prototype, we do not actually use the JPA to amplify an input signal. Instead, we use it to perform \emph{two-mode spontaneous downconversion}. This is done by causing the magnetic field to oscillate at the sum of two of the JPA's resonance frequencies without sending any signal into the JPA input port. In this case, the energy that is used to pump the magnetic field is converted into photons at the two resonance frequencies. The two beams thus created are entangled---in fact, they are the TMSV signals mentioned in Sec. \ref{subsec:TMSV}. This is another explanation for the ``vacuum'' part of ``two-mode squeezed vacuum'': we ``amplify'' the vacuum instead of some input signal. (One of the more remarkable results of quantum field theory is that, even in absolute vacuum, particles can fluctuate into and out of existence; in effect we are amplifying these fluctuations.)

\begin{figure}[t]
	\centerline{\includegraphics[width=0.8\columnwidth]{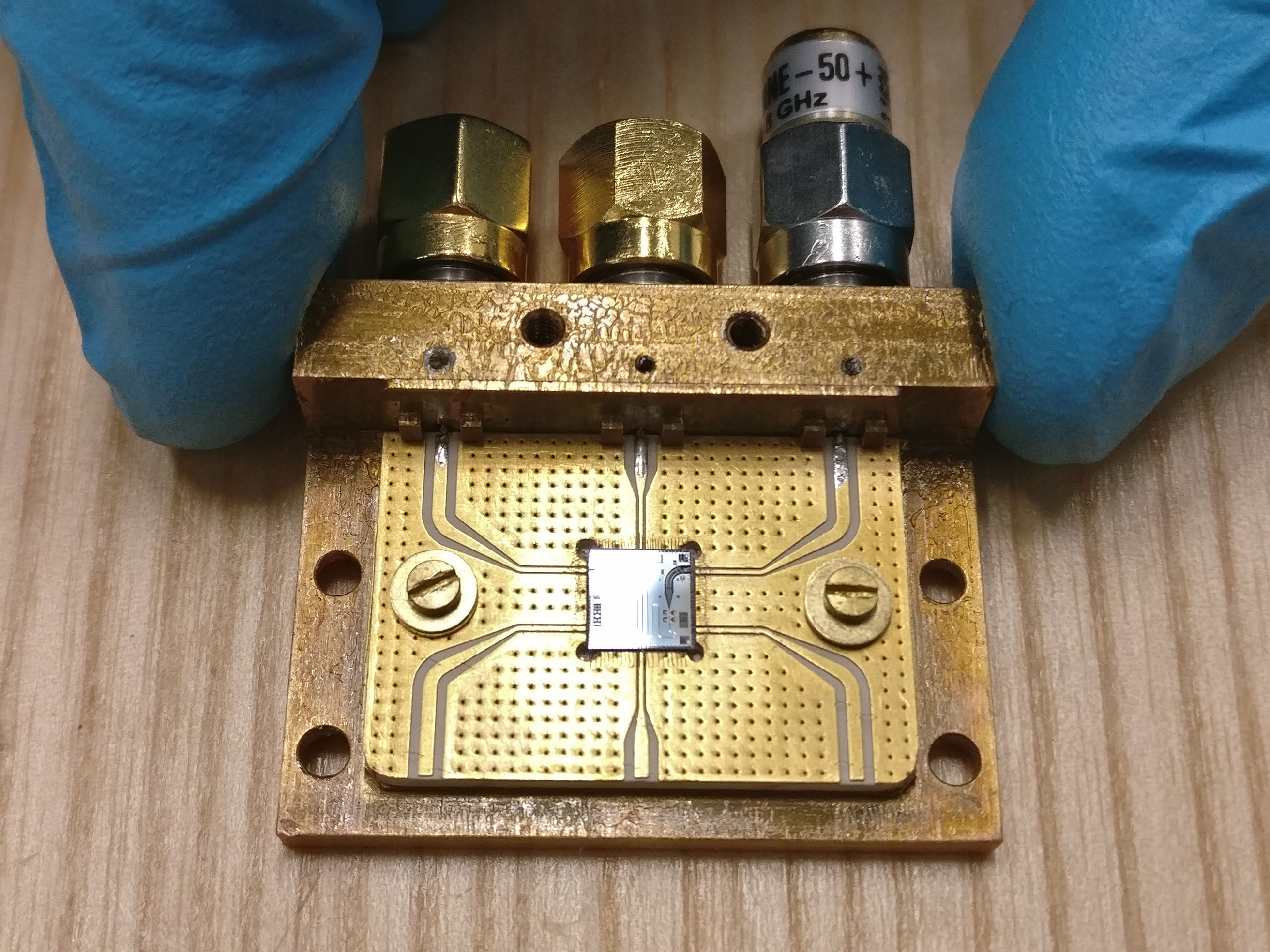}}
	\caption{Josephson parametric amplifier (JPA) mounted on a printed circuit board. This picture was reused from \cite{luong2018mqr}.}
	\label{fig:JPA_chip}
\end{figure}

The JPA we used in our quantum radar experiment was fabricated in-house at the University of Waterloo except for a final evaporation step, which was performed at the University of Syracuse. (The evaporation can now also be done in-house.) It consists of a quarter-wavelength resonator in the form of a 33 mm coplanar waveguide fabricated using thin-film aluminum. The waveguide is terminated at one end by a SQUID shunted to ground. At the other end, it is coupled to the input/output line through a small capacitance. The JPA is mounted on a printed circuit board that is approximately 3 cm long (Fig.\ \ref{fig:JPA_chip}). Apart from input and output ports for microwave signals, there is an input port for the oscillating magnetic field to which the SQUID is exposed. Just outside the casing holding the JPA, we have mounted a coil which produces a static magnetic field. (The total magnetic field is thus the sum of the static field produced by the coil and the oscillating field which is inputted into the chip.) 

It is not possible to operate a JPA at room temperature because it relies on the special behavior of superconducting materials. However, even if room-temperature superconducting materials were developed, it is unlikely that this would help. A JPA will not generate an entangled signal if the input has too much thermal noise. It is, after all, an amplifier (in fact, it is the ``best possible'' amplifier in that it adds the least amount of noise allowed by quantum physics), so one must be very careful as to what exactly is being amplified. Although they typically reside inside a cryogenic refrigerator, precautions have to be taken (in the form of the two circulators, for example) to prevent stray signals from entering the JPA's input port.

For the current experiment, the output signals of the JPA are at frequencies of 6.1445 GHz and 7.5376 GHz, these being two of the resonance frequencies after the SQUID is biased with a static magnetic field. (The TMN radar frequencies described in Sec.\ \ref{sec:classical_setup} were chosen to coincide with these values.) These frequencies can be tuned to some extent by changing the magnetic field strength.

\section{Signal Processing, Matched Filtering, and Detector Functions}
\label{sec:detectors}

As mentioned in Sec.\ \ref{sec:abstract_setup}, the operating principle behind our prototype QTMS radar requires us to perform matched filtering between the received signal and the signal that was recorded inside the system. We declare a detection when the output is higher than a given threshold. Because our transmit and receive horns are a fixed distance of 0.5 m away from each other, we only analyze the samples corresponding to the one-way path length of 0.5 m. 

In order to perform this matched filtering, we use the off-diagonal block matrix $R_{12}(\tau)$ in \eqref{eq:cov_blocks}, as mentioned in the discussion following that equation. When we compare \eqref{eq:cov_blocks} to the TMN radar covariance matrix \eqref{eq:classical_cov}, it appears that the parameter $\rho$ is the specific quantity we want to recover from our data. However, the sample covariance matrices we calculate from our experimental results do not fit the form of \eqref{eq:classical_cov} to infinite precision. (This covariance matrix is for the classical case, but as we have remarked earlier, the TMSV covariance matrix \eqref{eq:TMSV_cov} has the same form.) There are multiple detector functions which we can apply to our experimental data, all of which extract $\rho$ in the ideal case, but which may perform differently in real life. Moreover, there may be other functions that we can calculate which are not directly related to $\rho$, but which still serve to distinguish between signal and noise.

We also need to deal with the phase difference between the digitizers. Due to the limited phase coherence of the LOs in our digitizers when we lock them together, this phase drifts randomly over time. This corresponds to a drift in the value of $\phi$ in \eqref{eq:classical_cov} or \eqref{eq:TMSV_cov}. Because the phase is not important in this proof-of-concept experiment, we have preprocessed our measurement data in order to ``rotate'' it back to $\phi = 0$. This was done by calculating the sample covariance matrix for one second's worth of data, fitting a value of $\phi$ to the matrix, applying the appropriate rotation operator to the data to zero out the phase, and repeating this process for all of our collected data. Ideally, under this transformation the covariance matrix \eqref{eq:classical_cov} would reduce to
\begin{equation} \label{eq:classical_cov_rotated}
	\expval{xx^T} = 
	\begin{bmatrix}
		\sigma_1^2 & 0 & \rho \sigma_1\sigma_2 & 0 \\
		0 & \sigma_1^2 & 0 & -\rho \sigma_1\sigma_2 \\
		\rho \sigma_1\sigma_2 & 0 & \sigma_2^2 & 0 \\
		0 & -\rho \sigma_1\sigma_2 & 0 & \sigma_2^2
	\end{bmatrix} \!\! .
\end{equation}
If we were attempting to detect a target, its presence or absence would be indicated by whether the off-diagonal elements are zero or nonzero, respectively.

In this paper we examine the following detector functions:
\begin{center}
	\begin{enumerate}
	\setlength{\itemsep}{.5\baselineskip}
		\item \quad $\expval{I_1 I_2 - Q_1 Q_2}$
		\item \quad $\left| \expval{I_1 I_2 - Q_1 Q_2} \right|$
		\item \quad $\sqrt{\expval{I_1 I_2 - Q_1 Q_2}^{2} + \expval{I_1 Q_2 + Q_1 I_2}^2}$
		\item \quad $\sqrt{\expval{I_1 I_2}^2 + \expval{I_1 Q_2}^2 + \expval{Q_1 I_2}^2 + \expval{Q_1 Q_2}^2}$
		\item \quad $\operatorname{cov}\!\left(\! \sqrt{I_1^2 + Q_1^2}, \sqrt{I_2^2 + Q_2^2} \right)$
	\end{enumerate}
\end{center}
These detectors were, in part, inspired by expressions in the noise radar literature. See, for example, equations 15 and 16 of \cite{dawood2001roc} (though we use different notation). These functions are also motivated by the form of the TMN covariance matrix \eqref{eq:classical_cov} and the TMSV covariance matrix \eqref{eq:TMSV_cov}. If our sample covariance matrices were exactly of the form \eqref{eq:classical_cov}, the outputs of Detectors 1 to 4 would all be proportional to $\rho$.

Detector 5 discards the covariance between the two frequency modes and considers only the signal powers. In other words, this detector does not perform any matched filtering. Because of this, we expect that there should be no difference between the classical and quantum case when using this detector.

Note that Detectors 1 and 2 depend on the phase $\phi$, while Detectors 3--5 are independent of it. In our particular case we have preprocessed our data to set $\phi = 0$, but in other cases it may be desirable to omit the preprocessing.

In calculating these detectors, we do not need to perform the same type of careful calibration that is required to test for entanglement (as discussed in Section \ref{subsec:TMSV}). If that were necessary, it would have rendered the QTMS radar impractical.

We emphasize that, in terms of signal processing, we have treated the classical and quantum measurement results identically. The same mathematical formulas and computer code were used for both cases.

Note that this list of detector functions is not exhaustive, nor are any of the functions claimed to be optimal in any way.

\section{Results}
\label{sec:results}

\begin{figure*}[t!]
	\centering
	\subfloat[]{\includegraphics[width=.95\columnwidth]{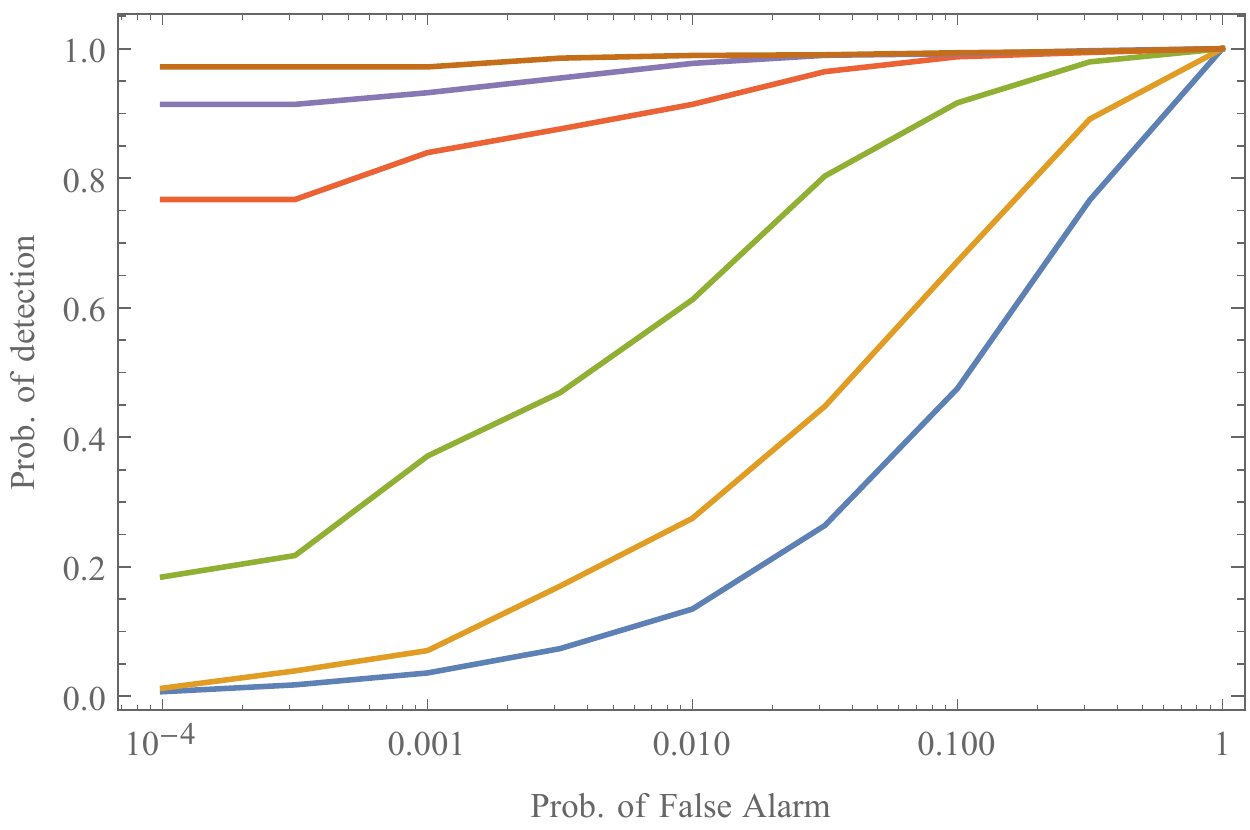}
		\label{subfig:ROC_quantum_det1}}
	\hfil
	\subfloat[]{\includegraphics[width=.95\columnwidth]{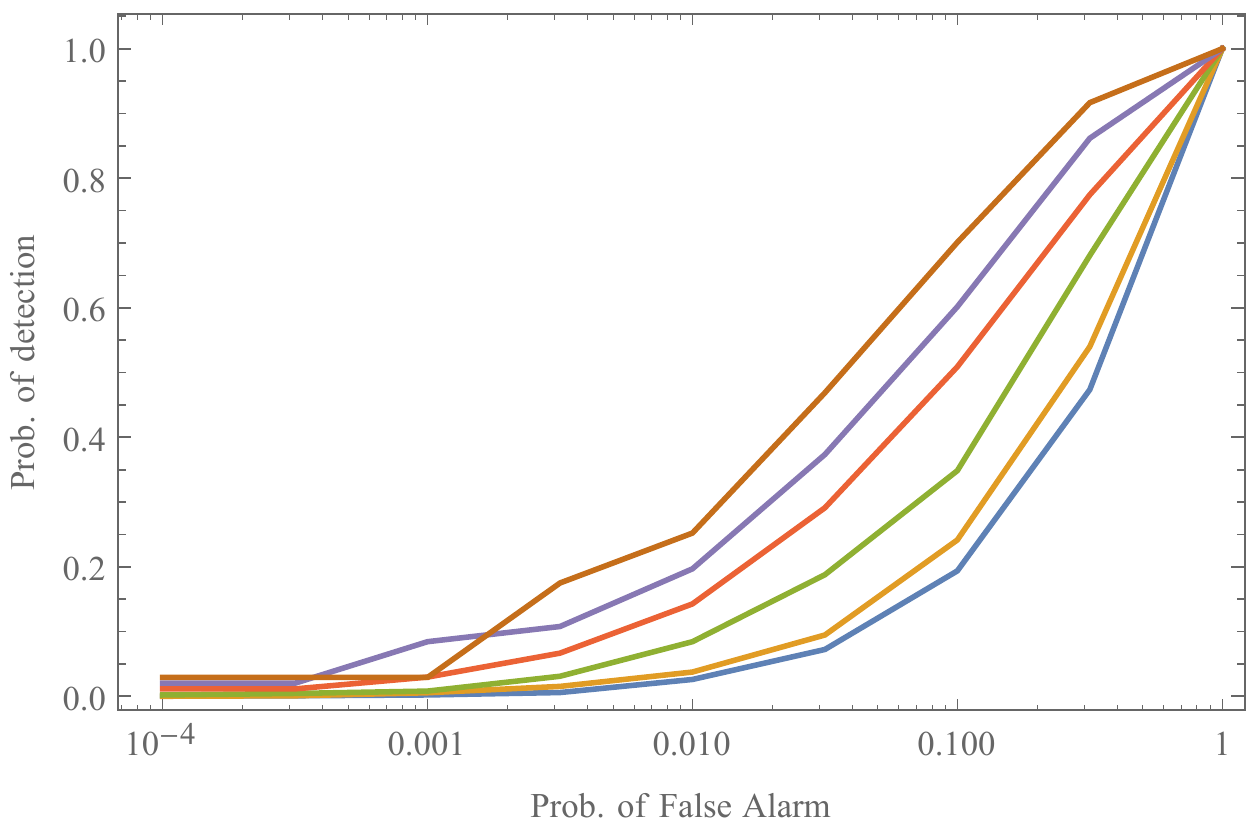}
		\label{subfig:ROC_classical_det1}}
	\caption{Receiver operator characteristic (ROC) curves for (a) the QTMS radar and (b) the TMN radar, both operating at a power level of roughly \textminus82 dBm, using Detector 1. In each plot, the curves (from bottom to top) are the result of integrating 5000, 10,000, 25,000, 50,000, 75,000, and 100,000 samples, respectively, corresponding to integration times of 5 ms, 10 ms, 25 ms, 50 ms, 75 ms, and 0.1 s.}
	\label{fig:ROC_quantum_classical_det1}
\end{figure*}

\begin{figure}[t]
	\centerline{\includegraphics[width=.95\columnwidth]{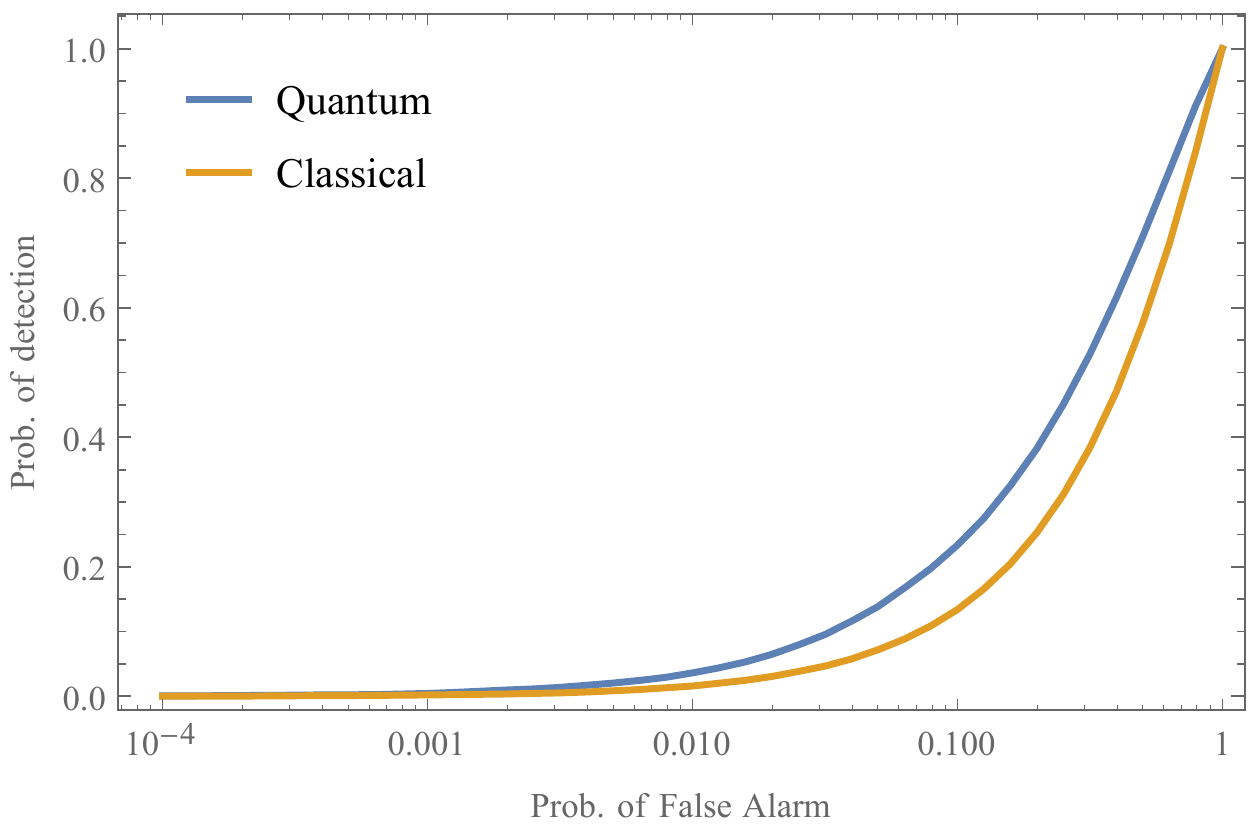}}
	\caption{Receiver operator characteristic (ROC) curves for the QTMS and TMN radars, both operating at a power level of roughly \textminus82 dBm, using Detector 1, with 1000 samples integrated.}
	\label{fig:ROC_det1_1k}
\end{figure}

For all the results in this section, the power level of the signal as it is being fed into the transmit horn is roughly \textminus82 dBm (as mentioned in Table \ref{table:parameters}). This is true for both the quantum and classical radars. Note that this power level is \emph{after amplification}. The signal power coming out of the JPA itself is lower: approximately \textminus145.43 dBm. These numbers correspond to approximately $5.7 \times 10^5$ photons per second before amplification (for the 7.5376 GHz signal), and $1.1 \times 10^{12}$ photons/s after amplification. Because the bandwidth is 1 MHz, the JPA emits roughly 0.57 photons/s per 1 Hz of bandwidth before amplification, and $1.1 \times 10^{6}$ (photons/s)/Hz after amplification. We can tune the power level of the JPA, but we found that operating the JPA at \textminus145.43 dBm seems to maximize the difference in performance between the quantum and classical setups. Note that the power levels required for our experiment are many orders of magnitude lower than the output power of typical noise radars like the one described in \cite{dawood2001roc}, which transmits at a power of 0 dBm (1 mW).

\subsection{QTMS vs. TMN Radar}

One of the main results of our experiment is shown in Fig.\ \ref{fig:ROC_quantum_classical_det1}. It shows receiver operating characteristic (ROC) curves for Detector 1 when we integrate 5000, 10,000, 25,000, 50,000, 75,000, and 100,000 samples. Because the sampling frequency of our setup is 1 MHz, these correspond to integration times of 5 ms, 10 ms, 25 ms, 50 ms, 75 ms, and 0.1 s respectively. Only a glance is needed to see that the QTMS radar (Fig.\ \ref{subfig:ROC_quantum_det1}) is markedly superior to the TMN radar (Fig.\ \ref{subfig:ROC_classical_det1}).

These plots also show that, as we increase the number of samples integrated, the difference in performance between the two radars becomes larger. For example, when the probability of false alarm ($p_\text{FA}$) is 0.001, the probability of detection ($p_\text{D}$) for both cases is almost the same when we integrate 5000 samples. But once we increase to 100,000 samples, the difference is dramatic: $p_\text{D} \approx 0.95$ for the quantum radar while $p_\text{D} \approx 0.05$ for the classical radar.

Fig.\ \ref{fig:ROC_det1_1k} shows ROC curves for both the quantum and classical radars for Detector 1 when we integrate only 1000 samples. In this case the improvement of the quantum radar over the classical one, while still present, is very small.

\begin{figure*}[p]
	\centering
	\subfloat[]{\includegraphics[width=.95\columnwidth]{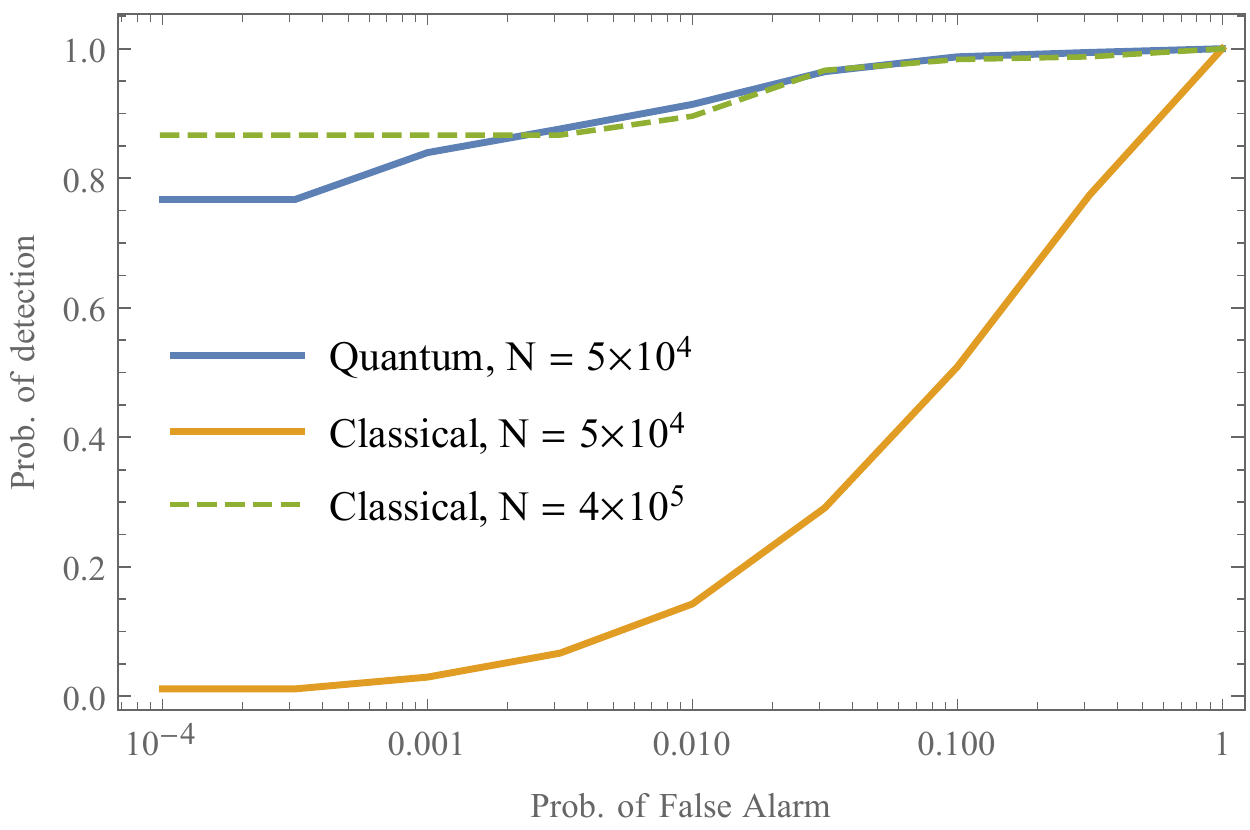}
		\label{subfig:ROC_integration_gain_det1_50k}}
	\hfil
	\subfloat[]{\includegraphics[width=.95\columnwidth]{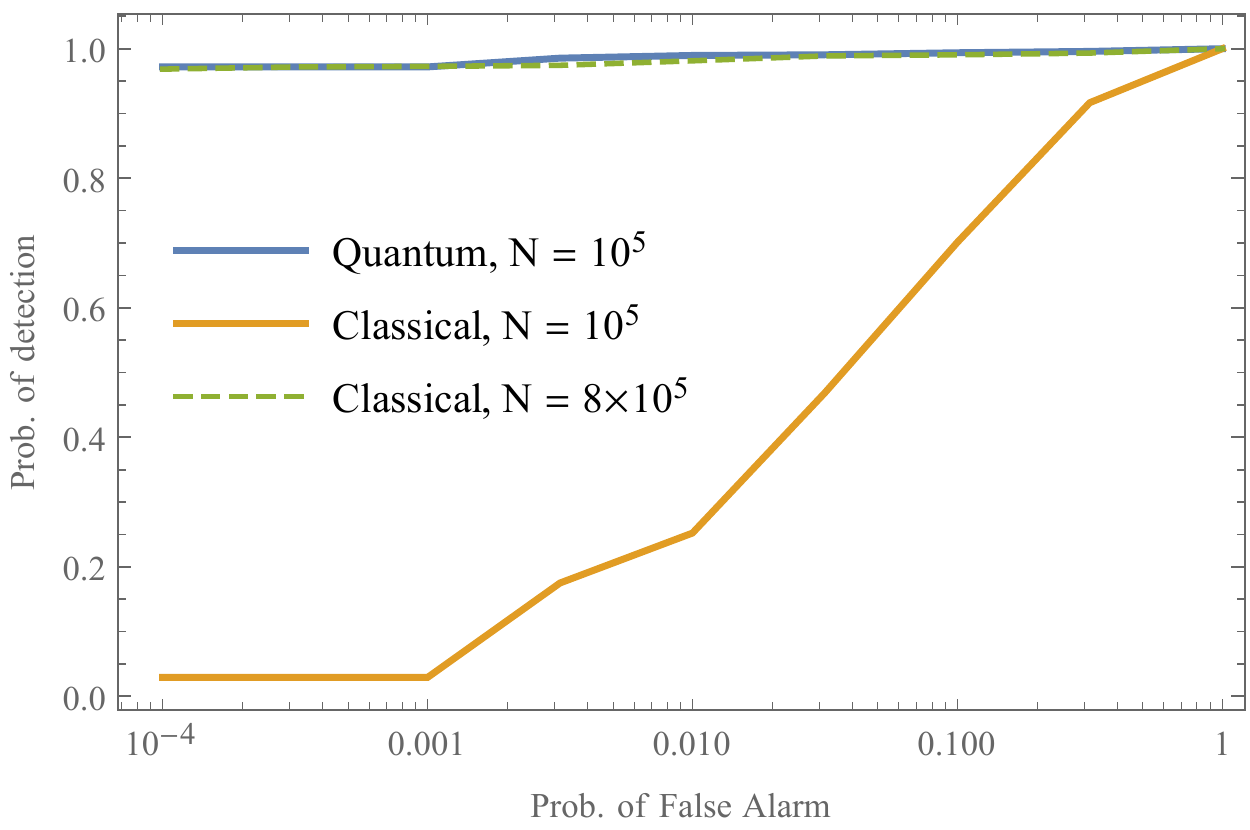}
		\label{subfig:ROC_integration_gain_det1_100k}}
	\caption{Solid lines: receiver operator characteristic (ROC) curves for our QTMS and TMN radars, using Detector 1, when integrating (a) 50,000 samples and (b) 100,000 samples. Dashed lines: ROC curves for the TMN radar when integrating (a) 400,000 samples and (b) 800,000 samples, showing that the TMN radar requires 8 times more samples to achieve results comparable to our QTMS radar.}
	\label{fig:ROC_integration_gain_det1}
\end{figure*}

One way to quantify the superiority of the QTMS radar over the TMN radar is displayed in Fig.\ \ref{fig:ROC_integration_gain_det1}. Again using Detector 1, we have plotted ROC curves for both radars when integrating 50,000 samples (Fig.\ \ref{subfig:ROC_integration_gain_det1_50k}) and 100,000 samples (Fig.\ \ref{subfig:ROC_integration_gain_det1_100k}). In addition, we have plotted ROC curves for the TMN radar when integrating 400,000 and 800,000 samples. Figure 12 illustrates that \textbf{the quantum radar achieves the same performance as the classical radar while reducing the number of samples integrated (hence the integration time) by a factor of eight}.

Although we show this only for 50,000 samples and 100,000 samples, the factor of 8 seems to hold for other values as well. In particular, we were also able to verify this when integrating 1000 samples (not shown here).

Note that our plots only go down to a probability of false alarm ($p_\text{FA}$) of $10^{-4}$ because we did not collect enough data to go lower. Nevertheless, the quantum advantage is clear.

We acknowledge that the performance of our TMN radar is worse than most practical radars and that the number of samples integrated is somewhat large. This can be explained by a number of factors: the simplicity of the setup (see Sec.\ \ref{sec:classical_setup}), the presence of multipath (see Fig.\ \ref{fig:horns_facing}), operation in the near field, and the extremely low power levels. The important point here is that we are performing an apples-to-apples comparison: same frequencies, same horns, same detection setup, same signal processing.

\begin{figure*}[p]
	\centering
	\subfloat[]{\includegraphics[width=.95\columnwidth]{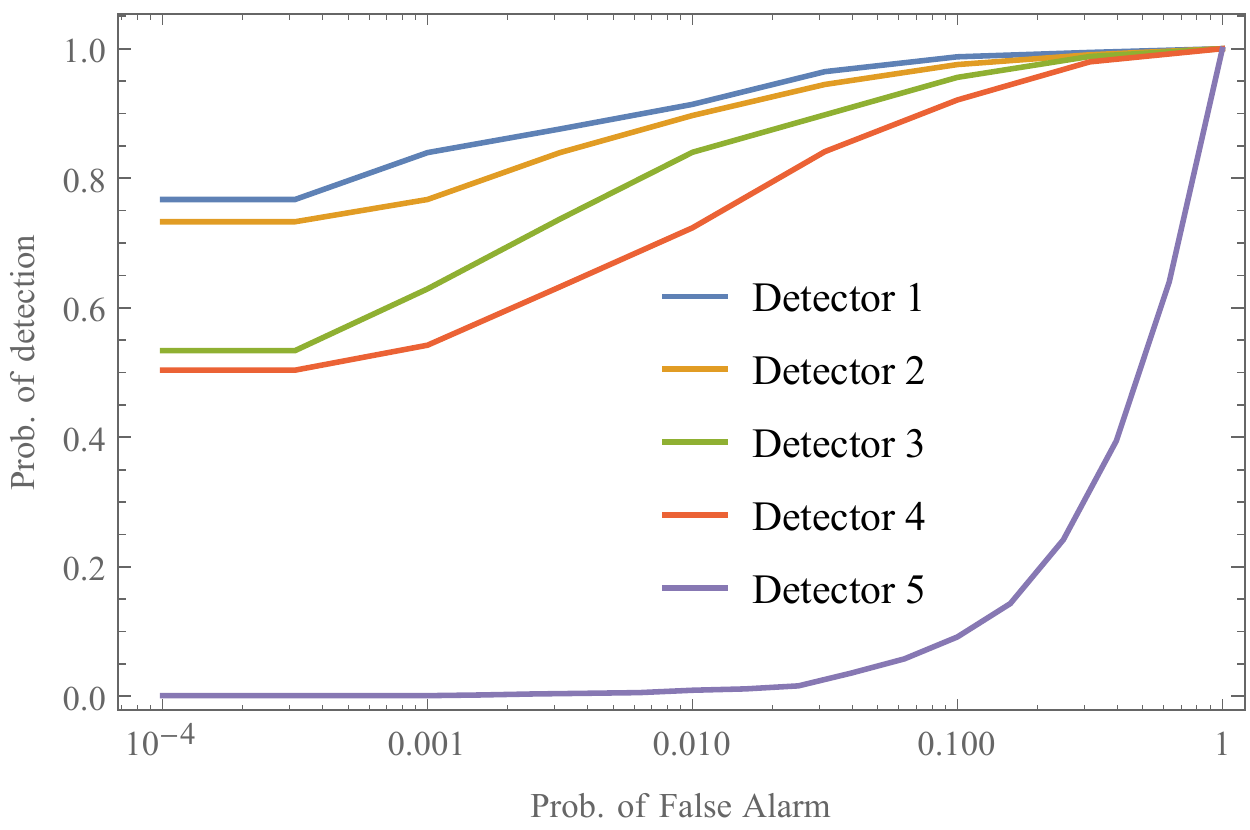}
		\label{subfig:ROC_quantum_detectors_50k}}
	\hfil
	\subfloat[]{\includegraphics[width=.95\columnwidth]{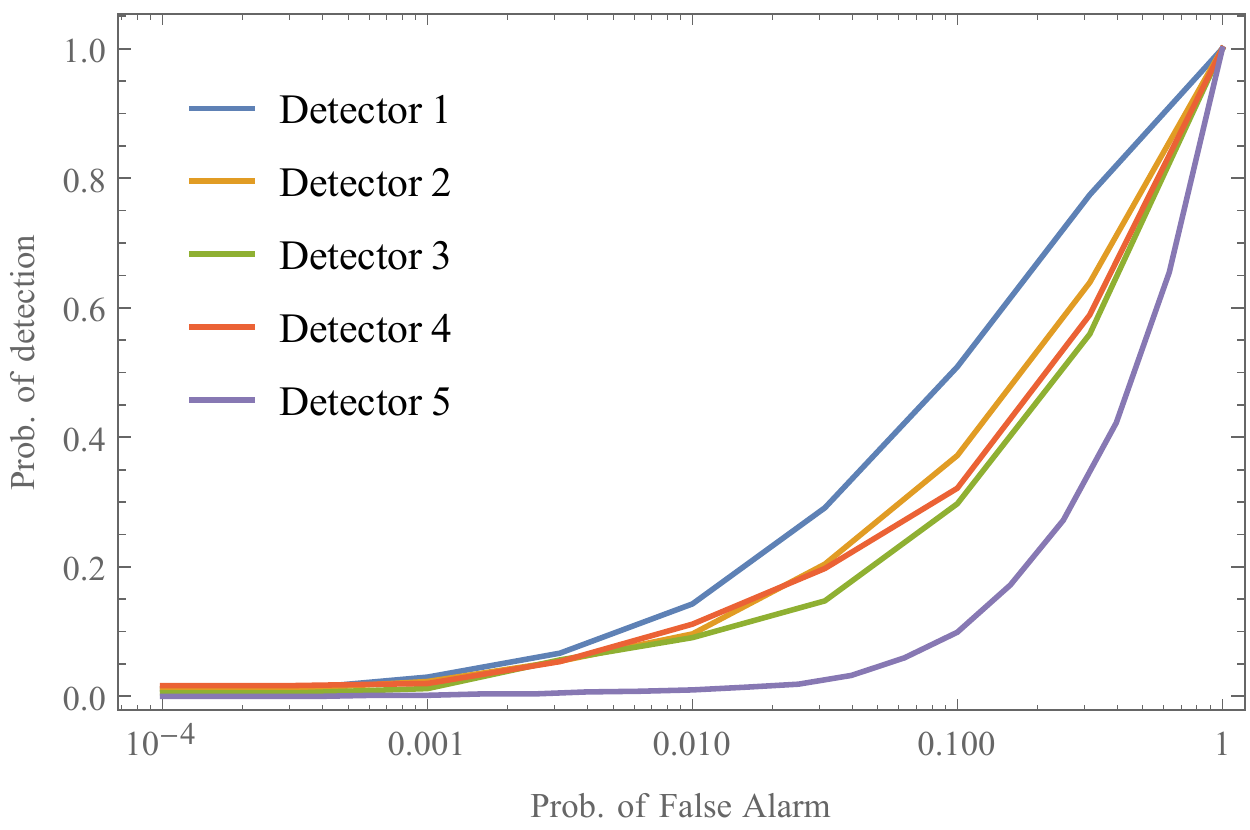}
		\label{subfig:ROC_classical_detectors_50k}}
	\hfil
	\subfloat[]{\includegraphics[width=.95\columnwidth]{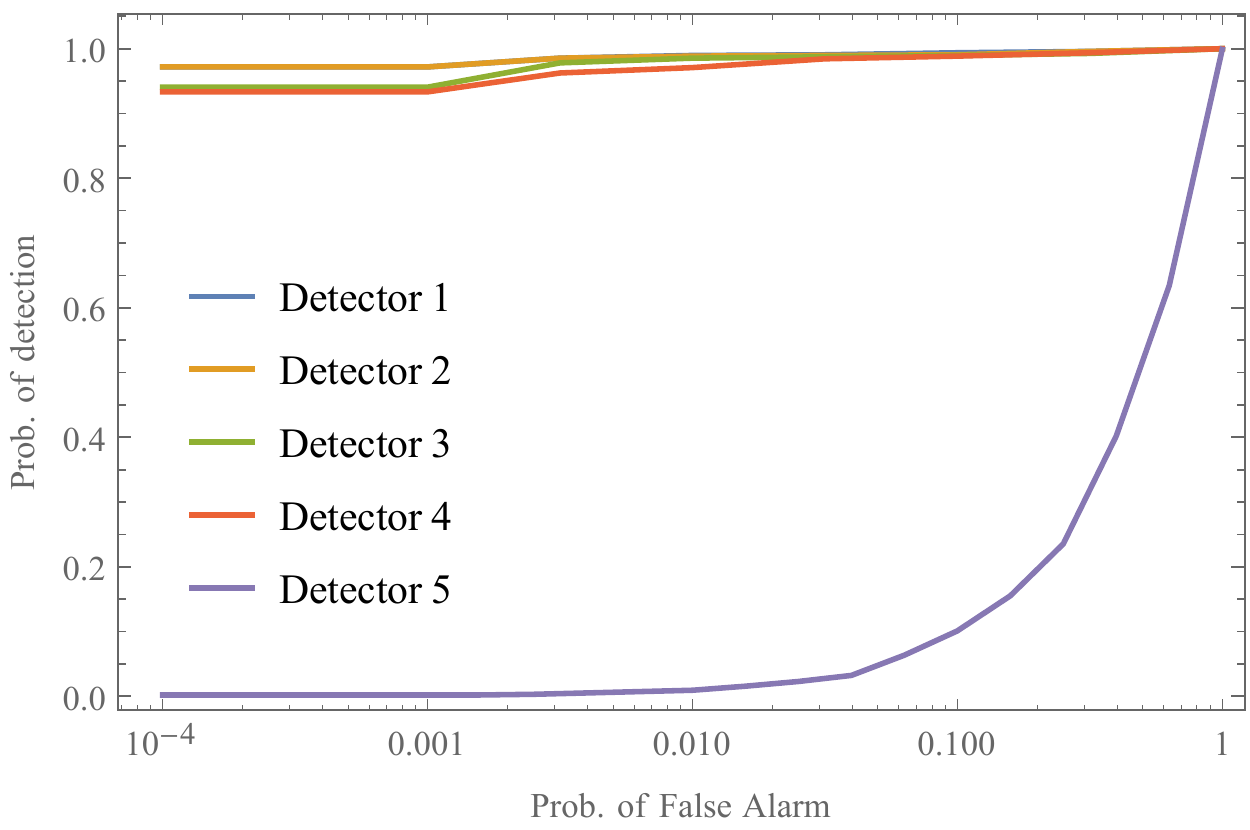}
		\label{subfig:ROC_quantum_detectors_100k}}
	\hfil
	\subfloat[]{\includegraphics[width=.95\columnwidth]{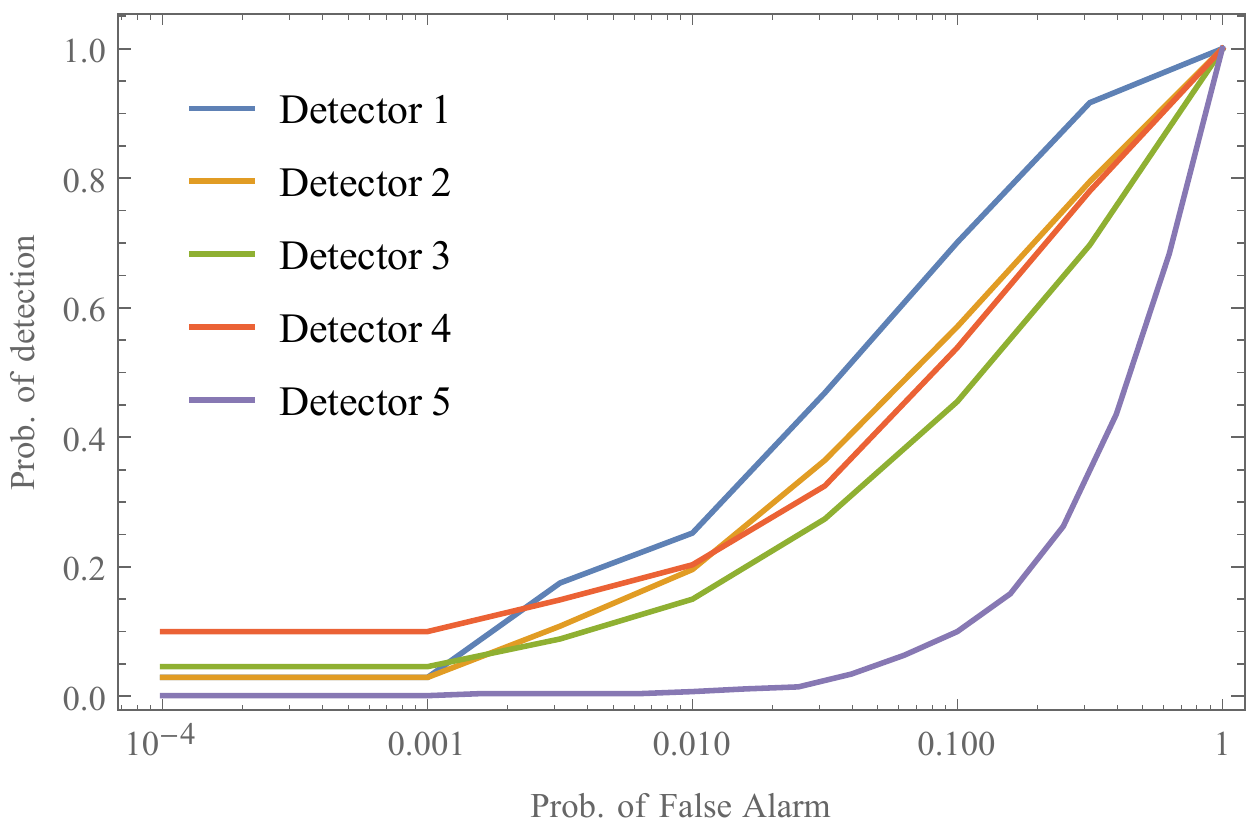}
		\label{subfig:ROC_classical_detectors_100k}}
	\caption{Comparison of detector function performance. (a) QTMS radar, 50,000 samples integrated. (b) TMN radar, 50,000 samples integrated. (c) QTMS radar, 100,000 samples integrated. (d) TMN radar, 100,000 samples integrated. Note that in (c), the curves for Detectors 1 and 2 are nearly identical, as are Detectors 3 and 4.}
	\label{fig:ROC_quantum_classical_detectors}
\end{figure*}

\begin{figure}[b]
	\centerline{\includegraphics[width=.95\columnwidth]{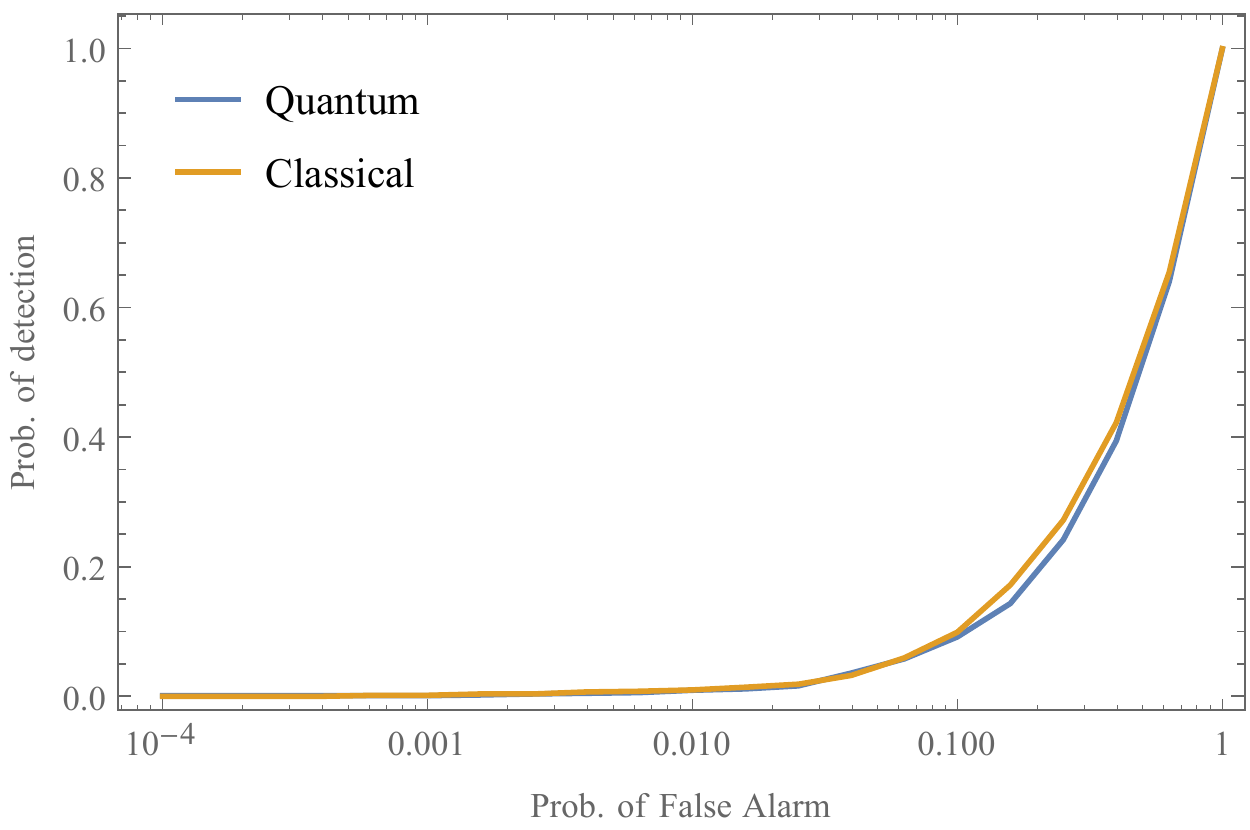}}
	\caption{Receiver operator characteristic (ROC) curves for our QTMS and TMN radars, both operating at a power level of roughly \textminus82 dBm, using Detector 5, with 50,000 samples integrated. Note that there is virtually no difference between the two curves.}
	\label{fig:ROC_det5_50k}
\end{figure}

\subsection{Comparison of Detector Functions}

In the previous subsection, we used only Detector 1 to generate our plots. The reason for this is because, of all five detectors, Detector 1 was found to be the best. This can be seen in Fig.\ \ref{fig:ROC_quantum_classical_detectors}, which is a comparison of the five detectors for both the QTMS and TMN radars when integrating 50,000 and 100,000 samples.

As stated, Detector 1 appears to be the best of all the detectors. However, with the exception of Detector 5, the difference between them appears to diminish as more samples are integrated. We believe that this is due to the rotation preprocessing which was mentioned in the previous section. By rotating the data so that $\phi = 0$, we are ``concentrating'' all of the covariance between the received and recorded signals in $\expval{I_1 I_2}$ and $\expval{Q_1 Q_2}$. This is evident by setting $\phi = 0$ in \eqref{eq:classical_cov} or \eqref{eq:TMSV_cov}. Therefore, any residual covariance in $\expval{I_1 Q_2}$ and $\expval{Q_1 I_2}$ is most likely just noise. Detectors 1 and 2 do not include contributions from this noise, which is why they are superior to Detectors 3 and 4. However, as more samples are integrated, the effect of the noise is suppressed. This explains why the difference between Detectors 1 and 2 vs.\ Detectors 3 and 4 diminishes as the number of samples increases.

We have seen that Detector 5 behaves differently: it is by far the worst of all the detectors, and the corresponding ROC curves do not appear to change with integration time. Moreover, Fig.\ \ref{fig:ROC_det5_50k} shows that there is no difference in performance between the quantum and classical radars when we use Detector 5. The reason for this is simple: Detector 5 looks only at power correlations between the two correlated signals, completely discarding the amplitude correlations. (Indeed, we cannot be said to be performing matched filtering when using this detector.) This shows that, in the scheme outlined in Sec. \ref{sec:abstract_setup}, amplitude correlations between the noise signals are crucial. Figure \ref{fig:ROC_det5_50k} also confirms that the power levels being generated by both the quantum and classical signal generators are the same, so we really are performing an apples-to-apples comparison.

\subsection{Comparison with Analytic Expressions for a Conventional Noise Radar}

\begin{figure*}[t!]
	\centering
	\subfloat[]{\includegraphics[width=.95\columnwidth]{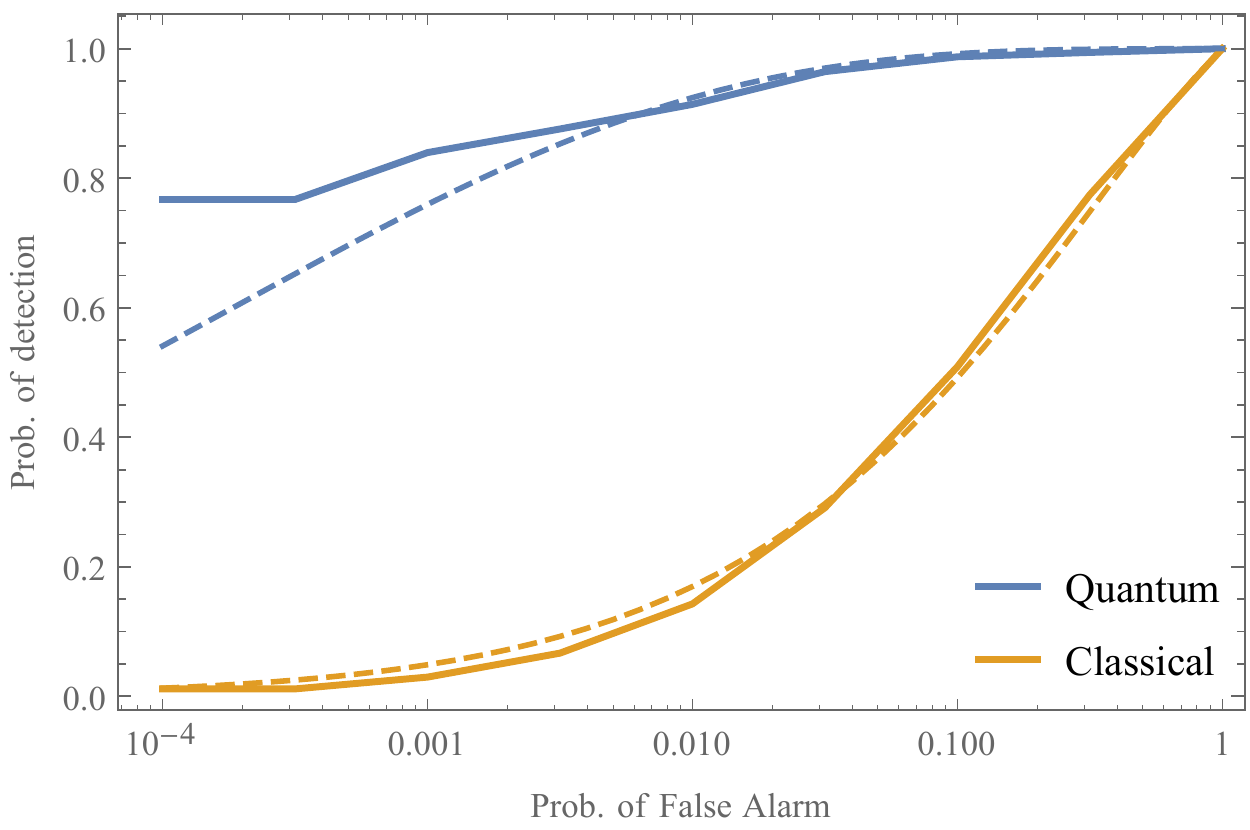}
		\label{subfig:ROC_det1_approx_50k}}
	\subfloat[]{\includegraphics[width=.95\columnwidth]{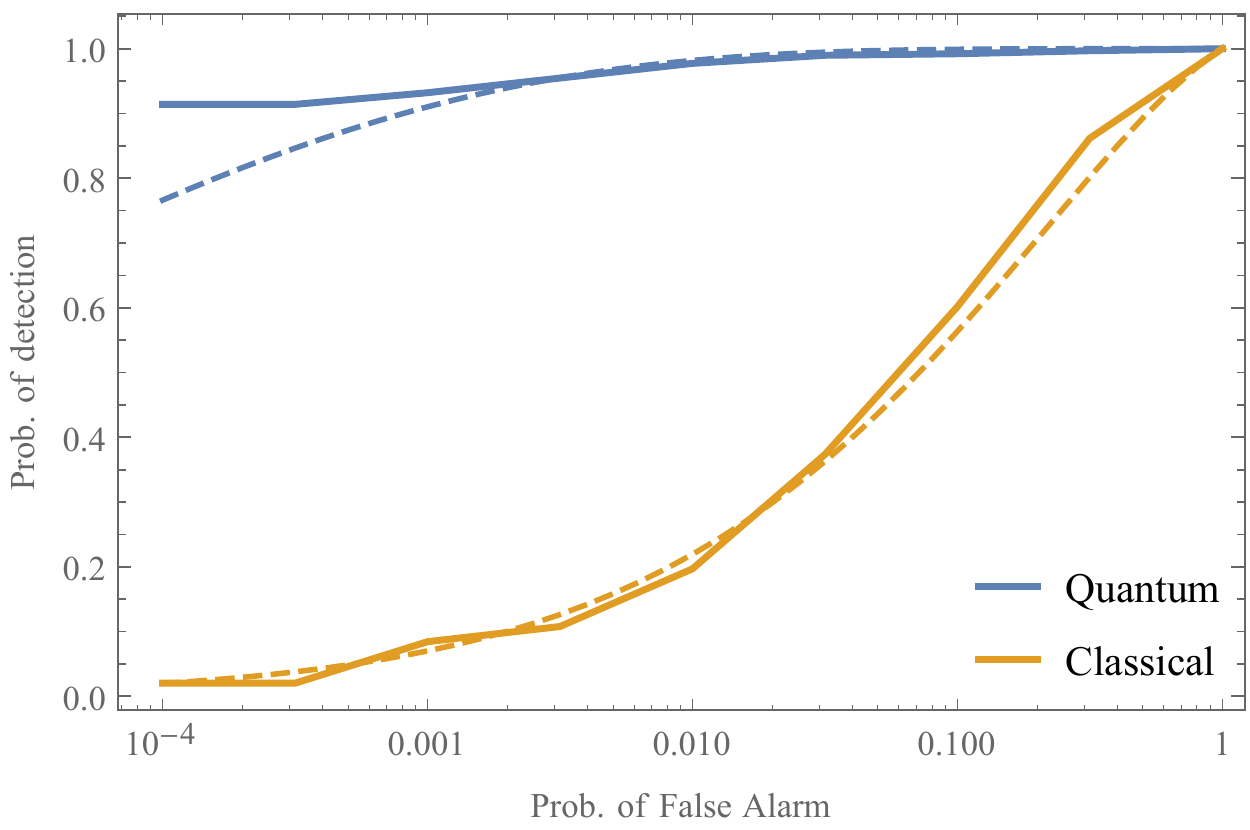}
		\label{subfig:ROC_det1_approx_75k}}
	\caption{Solid lines: receiver operator characteristic (ROC) curves for our QTMS and TMN radars, using Detector 1, when integrating (a) 50,000 samples and (b) 75,000 samples. Dashed lines: approximations to these ROC curves using analytical formulas in \cite{dawood2001roc}.}
	\label{fig:ROC_det1_approx}
\end{figure*}

In \cite{dawood2001roc}, Dawood and Narayanan analyzed a coherent ultrawideband random noise radar and derived analytic expressions for the cumulative distribution function of its detector output. This is a conventional noise radar which differs from our TMN radar in some respects, but there are enough similarities that we can attempt to fit our results to their mathematical formulas.

For Dawood and Narayanan's noise radar, the probability that the detector output $Z$ exceeds a threshold $T$ was calculated to be
\begin{equation}\label{eq:dawood_CDF}
\begin{split}
	p(Z > T) &= \frac{2^{N+1}}{(N-1)!} \tilde{T}^N  \\
	&\phantom{=} \times \sum_{m=0}^{\infty} \rho^m K_{N+m}\! \left( \! \frac{4\tilde{T}}{1-\rho^2} \! \right) I_m\! \left( \! \frac{4\rho\tilde{T}}{1-\rho^2} \! \right)
\end{split}
\end{equation}
where $\tilde{T}$ is the normalized threshold $T/(\sigma_1\sigma_2)$, $N$ is the number of samples integrated, $K_{N+m}$ is the modified Bessel function of the second kind of order $N+m$, and $I_m$ is the modified Bessel function of the first kind of order $m$. As in \eqref{eq:sigmarho}, $\rho$ is the correlation between the two measured signals while $\sigma_1^2$ and $\sigma_2^2$ are the signals' noise powers.

The probability of false alarm for this noise radar is obtained by taking the limit $\rho \to 0$:
\begin{equation}\label{eq:dawood_pFA}
	p_\text{FA} = \frac{2^{N+1}}{(N-1)!} \tilde{T}^N K_N(4\tilde{T}).
\end{equation}
From these expressions it is possible to calculate theoretical ROC curves.

Fig.\ \ref{fig:ROC_det1_approx} shows experimental ROC curves for our quantum and classical radars when integrating 50,000 samples (Fig.\ \ref{subfig:ROC_det1_approx_50k}) and 75,000 samples (Fig.\ \ref{subfig:ROC_det1_approx_75k}), together with approximations to these curves using \eqref{eq:dawood_CDF} and \eqref{eq:dawood_pFA}. We can see that the theoretical curves are generally a good fit to the experimental data. The fit for the classical TMN radar seems almost perfect, but the QTMS radar seems to deviate somewhat. This is expected because our quantum radar is not a conventional noise radar, but they are not radically different from one, either.

Due to numerical instabilities, we were unable to directly set $N = 50000$ or $N = 75000$ in \eqref{eq:dawood_CDF} and \eqref{eq:dawood_pFA}. Instead, we were forced to arbitrarily select a ``nominal'' $N$ and fit the curves by varying $\rho$. Obviously the values of $\rho$ thus obtained are meaningless in and of themselves, but the ratio $\rho_\text{quantum}/\rho_\text{classical}$ should still be meaningful. We have found that this ratio is about 2.4 when integrating 50,000 samples and 2.5 when integrating 75,000 samples. These ratios seem to remain constant when varying the nominal value of $N$, which increases our confidence that the ratio itself is meaningful.

Based on this rudimentary analysis, we can quantify the improvement of our QTMS radar over the TMN radar in terms of an improvement in the correlation coefficient $\rho$ of a comparable noise radar. Using this metric, we can say that our quantum radar generates correlations which are about \textbf{2.4 to 2.5 times stronger} than those of the TMN radar.

\begin{figure*}[p]
	\centering
	\subfloat[]{\includegraphics[width=.82\columnwidth]{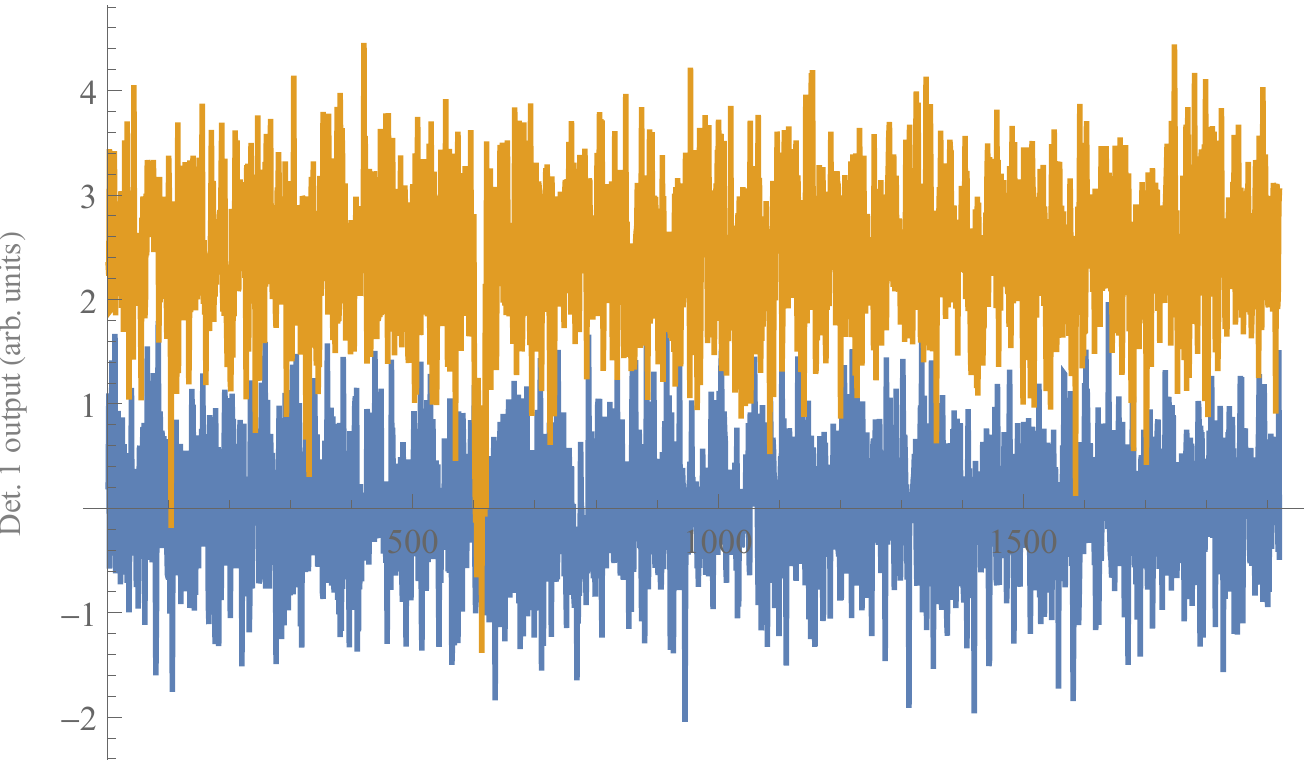}
		\label{subfig:timeseries_quantum_det1_50k}}
	\hfil
	\subfloat[]{\includegraphics[width=.82\columnwidth]{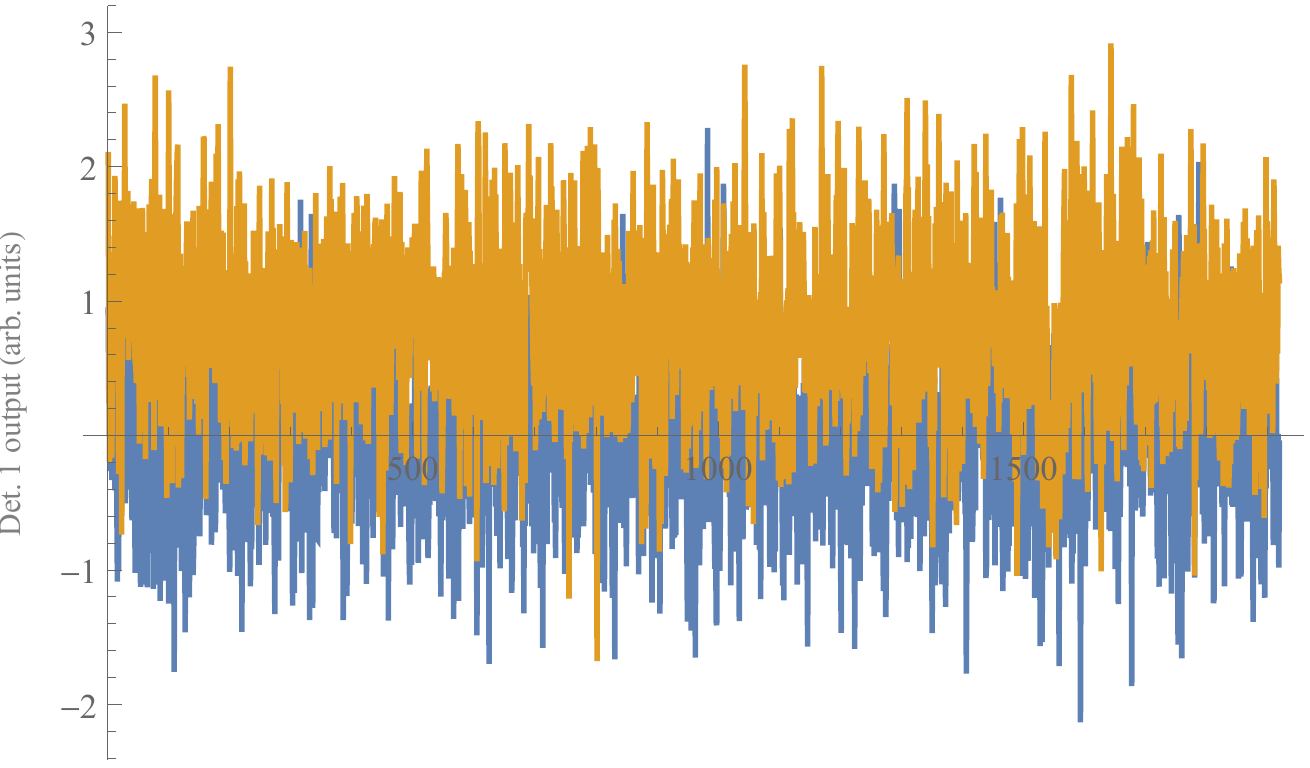}
		\label{subfig:timeseries_classical_det1_50k}}
	\hfil
	\subfloat[]{\includegraphics[width=.82\columnwidth]{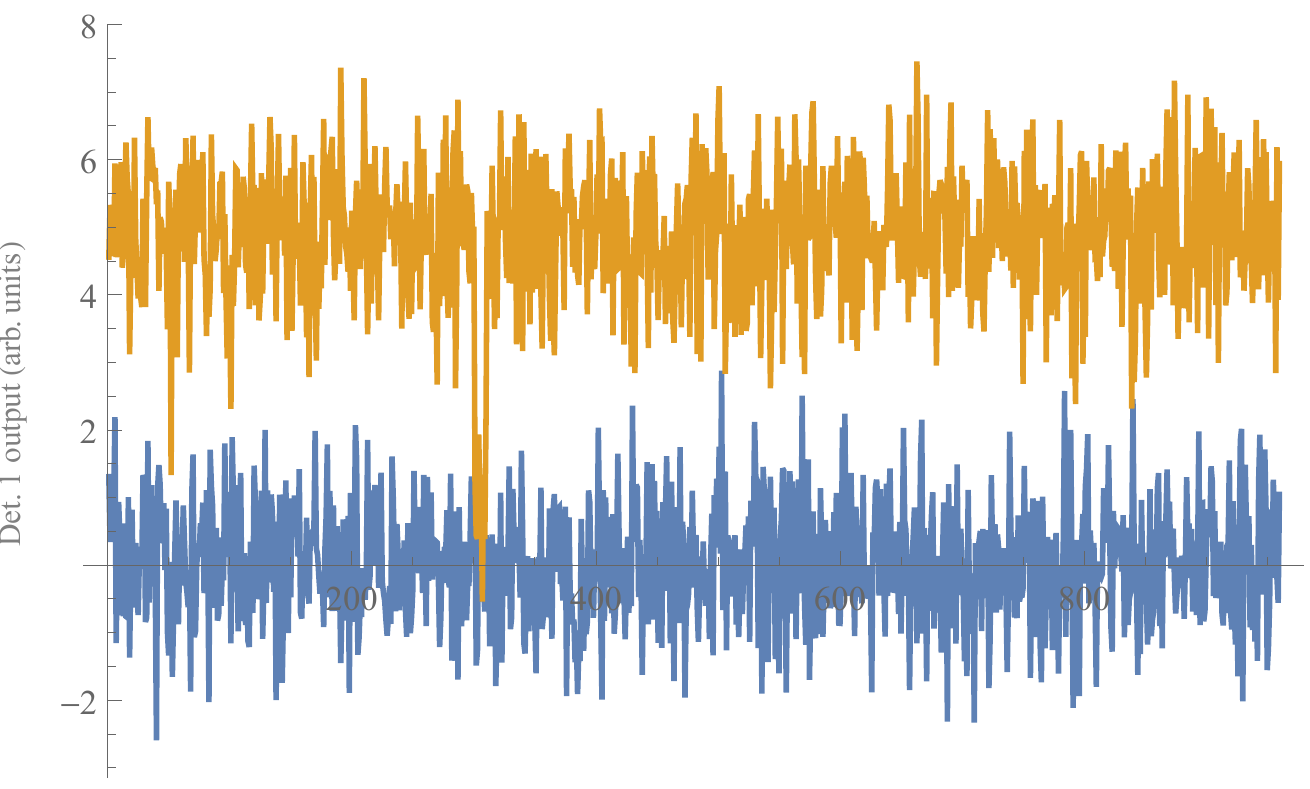}
		\label{subfig:timeseries_quantum_det1_100k}}
	\hfil
	\subfloat[]{\includegraphics[width=.82\columnwidth]{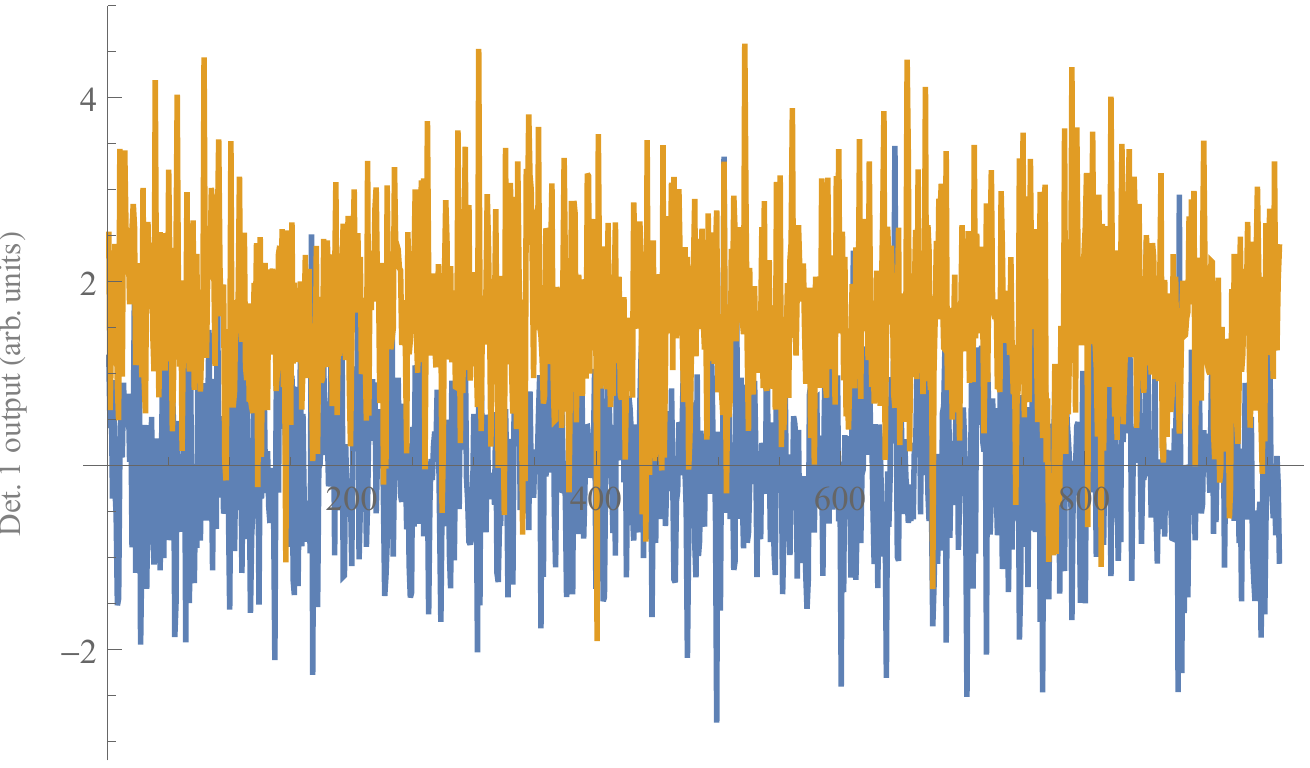}
		\label{subfig:timeseries_classical_det1_100k}}
	\caption{Time series of Detector 1 outputs (in arbitrary units). (a) QTMS radar, 50,000 samples integrated. (b) TMN radar, 50,000 samples integrated. (c) QTMS radar, 100,000 samples integrated. (d) TMN radar, 100,000 samples integrated. Blue (lower series): detections when the signal generator is turned off. Orange (upper series): detections when the signal generator is turned on.}
	\label{fig:timeseries_quantum_classical_det1}
\end{figure*}

\begin{figure*}[p]
	\centering
	\subfloat[]{\includegraphics[width=.82\columnwidth]{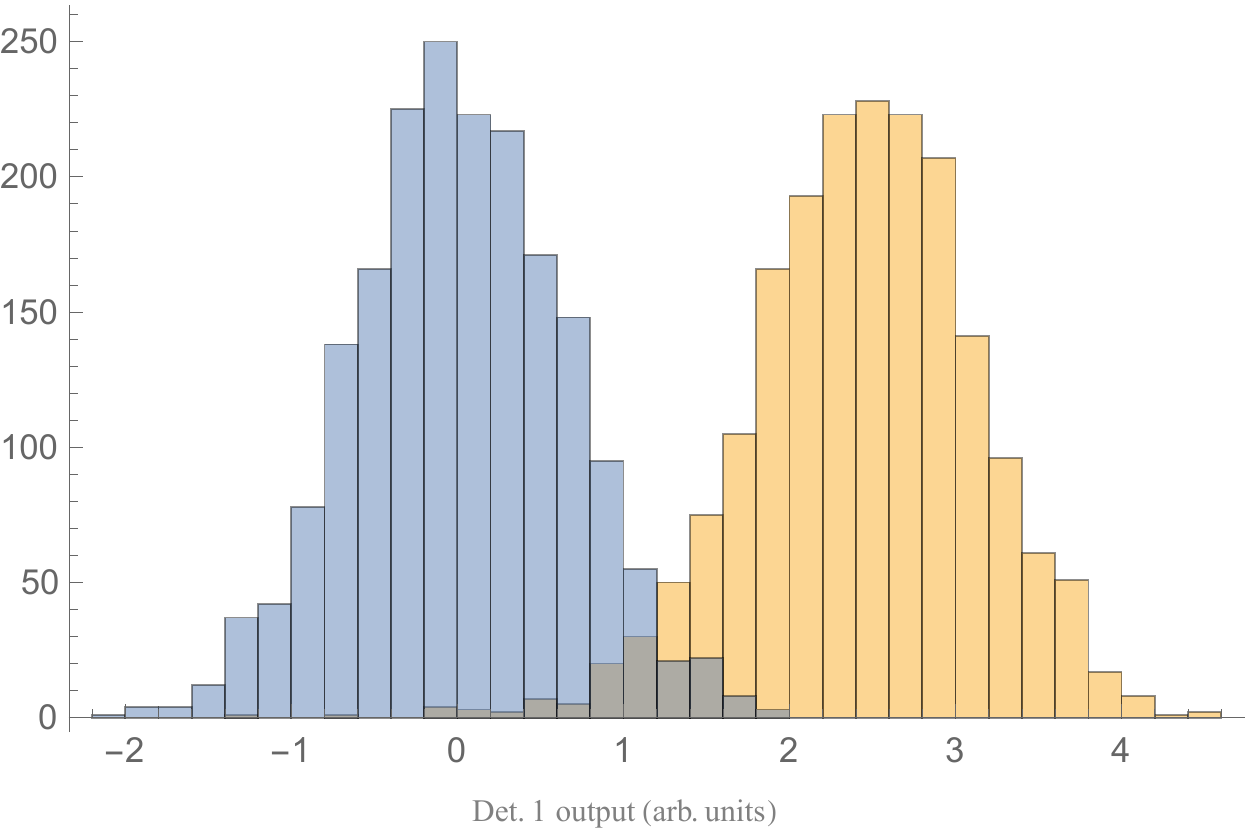}
		\label{subfig:histogram_quantum_det1_50k}}
	\hfil
	\subfloat[]{\includegraphics[width=.82\columnwidth]{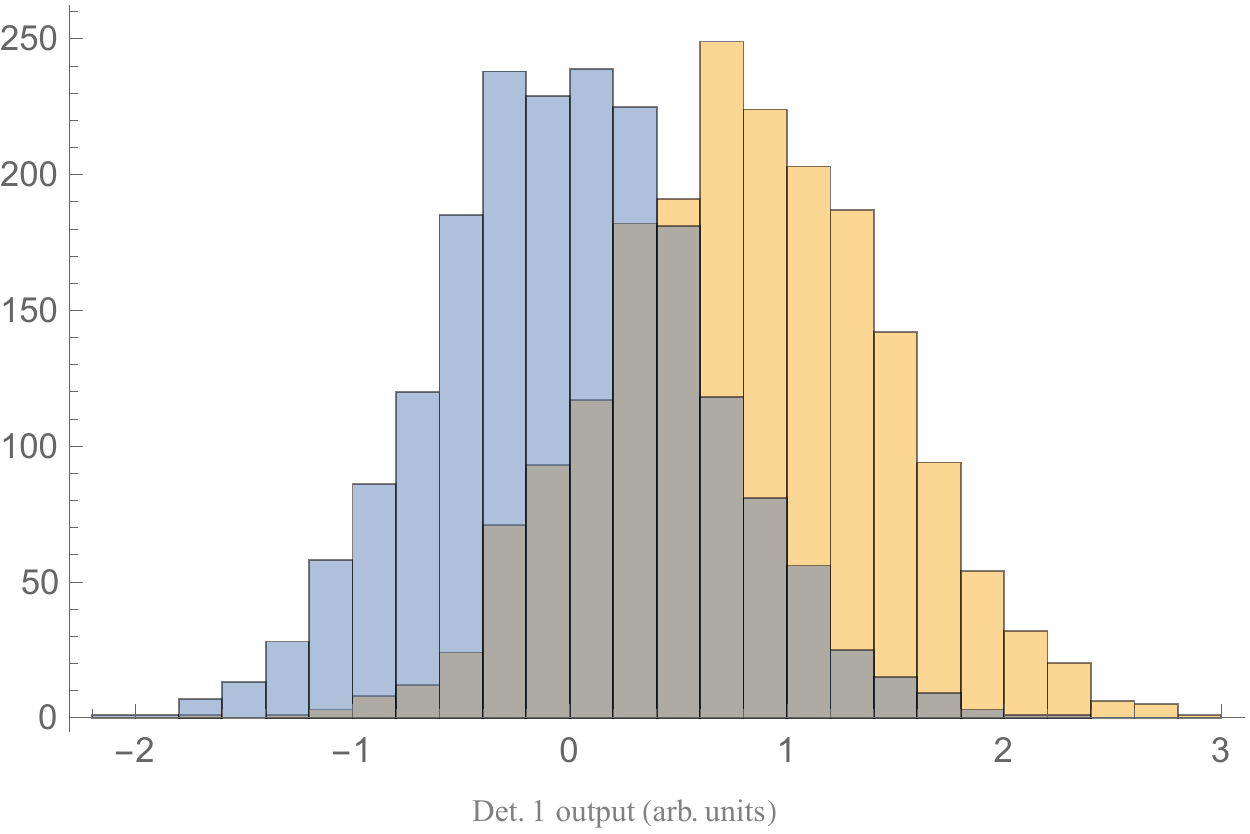}
		\label{subfig:histogram_classical_det1_50k}}
	\hfil
	\subfloat[]{\includegraphics[width=.82\columnwidth]{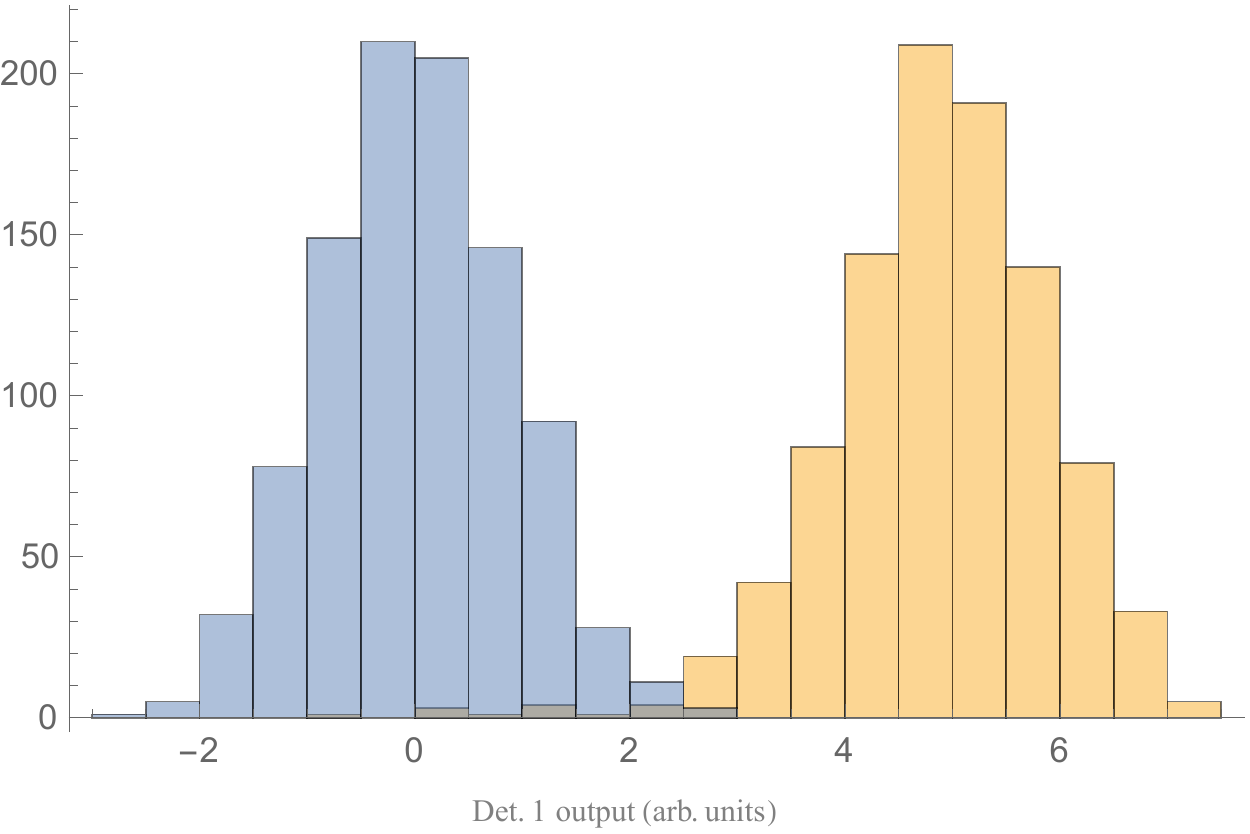}
		\label{subfig:histogram_quantum_det1_100k}}
	\hfil
	\subfloat[]{\includegraphics[width=.82\columnwidth]{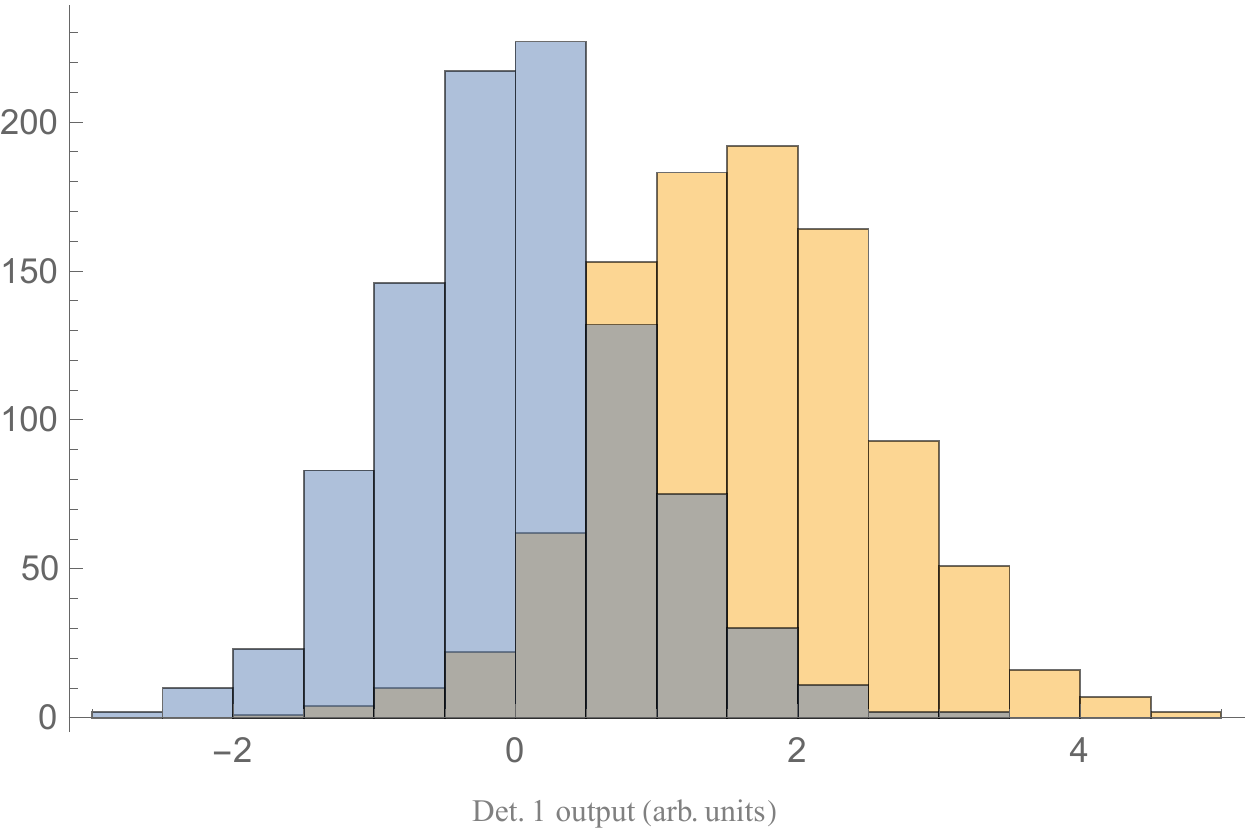}
		\label{subfig:histogram_classical_det1_100k}}
	\caption{Histograms showing the distribution of Detector 1 outputs (in arbitrary units). (a) QTMS radar, 50,000 samples integrated. (b) TMN radar, 50,000 samples integrated. (c) QTMS radar, 100,000 samples integrated. (d) TMN radar, 100,000 samples integrated. Blue (left peak): detections when the signal generator is turned off. Orange (right peak): detections when the signal generator is turned on.}
	\label{fig:histogram_quantum_classical_det1}
\end{figure*}

\subsection{Time Series and Histograms of Detector 1 Outputs}

In Fig.\ \ref{fig:timeseries_quantum_classical_det1}, we have plotted time series of Detector 1 outputs for our QTMS and TMN radars after integrating 50,000 samples and 100,000 samples. This was done both when the signal generators were turned on and when they were off. The plots show that the separation between on and off data is quite marked for the quantum radar, but somewhat less so for the classical radar. This explains the superiority of our quantum radar over its classical counterpart. Looking at Fig. \ref{subfig:timeseries_quantum_det1_100k}, corresponding to the QTMS radar with 100,000 samples integrated, we observe that the on and off time series would practically be completely distinct if not for one particularly large fluctuation. (This fluctuation might be due to the experiment being disturbed somehow during the data collection process.)

Another way of looking at this data is to plot histograms which show the distribution of Detector 1 outputs. Again, we have done so for our quantum radar and the classical radar after integrating 50,000 samples and 100,000 samples; the results are shown in Fig.\ \ref{fig:histogram_quantum_classical_det1}. We see the same clear separation between on and off data for the quantum case; the classical histograms show significant overlap.

Incidentally, these plots corroborate the assertion in the previous subsection that the quantum radar generates correlations which are about 2.4 to 2.5 times stronger than those of the TMN radar. In the arbitrary units chosen for Fig.\ \ref{subfig:histogram_quantum_det1_50k}, the covariance (when the device is on) appears to be centered around 2.5, whereas in Fig.\ \ref{subfig:histogram_classical_det1_50k} it appears to be centered around 1. The ratio of these two values is 2.5, which is consistent with the above statement. Similar statements hold when examining Figs.\ \ref{subfig:histogram_quantum_det1_100k} and \ref{subfig:histogram_classical_det1_100k}.

\section{Conclusion}

In our QTMS radar prototype, we have demonstrated all of the ingredients needed in a quantum-enhanced radar transmitter. In particular, we succeeded in transmitting a signal through free space which, while unentangled, contains strong correlations originating from an entangled signal generator which emits two-mode squeezed vacuum (TMSV) signals. All this was done entirely in the microwave regime, without downconversion from optical frequencies (in which the production of entangled signals is often easier). We were able to detect the transmitted signal at powers as low as \textminus82 dBm.

As part of our analysis, we compared our quantum radar with a classical radar which generates signals by mixing a carrier with Gaussian noise. It retains an architecture which is parallel to the quantum-enhanced radar architecture to support the thesis that, unless the two signals are entangled, the classical noise is correlated but the quantum noise is not. When the signals are entangled, the performance improvement is due to the correlation of the quantum noise. Thus we are conducting an apples-to-apples comparison. Under these conditions, we have found that there is a distinct quantum enhancement. In particular, \textbf{the QTMS radar reduces the integration time by a factor of eight} relative to the TMN radar.

The quantum illumination literature states that in the ideal case, quantum illumination exhibits an improvement of up to 6 dB in the error exponent for target detection (depending on the receiver) \cite{tan2008quantum}. We have deliberately avoided comparing our work to numbers like this because the stated improvements are over a putative ``optimal'' classical radar. Practically speaking, there is no such thing as a uniquely optimum radar: different radars are better under different criteria. We have instead compared our QTMS radar to an actual, physical system which we have built in a lab. Moreover, as discussed in Sec.\ \ref{sec:intro}, we are not performing traditional quantum illumination. Finally, we prefer not to make comparisons on the basis of this highly abstract error exponent. (For more on this point, see \cite{balaji2018snake}.) Our analysis, based on ROC curves, translates more directly to real-world performance.

There are four caveats which we wish to make about the work described in this paper:
\begin{enumerate}
	\item We have implemented only a one-way ranging scheme instead of reflecting the signal off a target. In future work we intend to attempt the detection of various targets at different ranges.
	\item The signals emerging from the refrigerator are not entangled, although they were entangled \emph{at the JPA source}. This leaves open the objection that our results could be reproduced simply by building a better classical system---in principle we could leave quantum mechanics out of the discussion entirely. In future work, we intend to remove the amplifier chain and extract the entangled quantum signal directly. If we succeed in doing so, the resulting transmitted signal would be \emph{theoretically impossible} to reproduce without an entangled source.
	\item We emphasize that we have not attempted to create any sort of ``optimal'' quantum radar. There is undoubtedly much room for improvement. We cannot deny that certain portions of our experimental setup were very crude. On the transmit side, the generated signals were narrowband. On the receive side, we had only a horn, an amplifier, and a digitizer. What's worse, we used X-band horns with C-band frequencies, and these horns were simply taped to a desk! Yet we were able to show a large quantum gain under these circumstances.
	\item We have not yet shown that our QTMS radar can be used to generate high-resolution SAR imagery, array processing for direction-finding, space-time adaptive processing, etc. However, the connections drawn in this paper enable the exploration of such concepts in future work.
	\item Our prototype QTMS radar is not practical in the sense that it is not immediately field-deployable. It requires a cryogenic refrigerator, time to cool to operating temperatures, and large power requirements (SWaP). However, our results are of interest as they are independent of the chosen technological route. Moreover, they motivate the development of suitable quantum information science for the development of more practical and scalable QTMS radars.
\end{enumerate}

Apart from these considerations, there are many other possible improvements to our scheme. For example, there are many schemes for generating various types of entangled signals at various frequency bands and bandwidths. On the signal processing side, there may also be better detector functions or other techniques which may be employed. Much exploratory work remains to be done.

TMSV is only one type of entangled signal; it falls within the class of bipartite, continuous-variable entangled signals. There are other types of quantum entanglement which do not fit this description, such as polarization entanglement, time-bin entanglement, N00N states, and multipartite entangled signals. (Bipartite entanglement is much easier to achieve than multipartite entanglement, which is required in most quantum computing applications.) Some of these have already been shown to be applicable to quantum radar. N00N states are used in interferometric quantum radar \cite{dowling2008n00n,lanzagorta2011quantum}. Another proposal uses photon-subtracted two-mode squeezed states (PSTMSS) \cite{zhang2014qi}; its authors show that ``quantum illumination with PSTMSS appreciably outperforms its classic correspondence in both low- and high-noise operating regimes, extending the regimes in which quantum illumination is optimal for target detection.'' It is not unreasonable to think that other types of entangled signals will be found to have applications to radar.

With the experiment described in this paper, it is no longer possible to say that microwave quantum radar is impossible or infeasible. Moreover, the results presented in Sec.\ \ref{sec:results} rule out the contention that the benefit of a quantum radar is merely marginal. On the contrary, we have demonstrated a significant gain over a similar classical system, including the possibility of reducing integration times dramatically. We hope that our results will spur further innovation in the novel and exciting field of quantum radar.

\section*{Acknowledgement}

We would like to thank M. Fahmi, T. Laneve, K. L\'{e}vis, N. Reed, P. Reitano, E. Riseborough, and P. Sevigny for advice and helpful discussions on data acquisition. We also thank J. Bourassa for his advice and comments on an earlier draft of this paper. This work was supported by Defence Research and Development Canada. C.M.W., C.W.S.C., and A.M.V. acknowledge additional funding from NSERC of Canada, the Canadian Foundation for Innovation, the Ontario Ministry of Research and Innovation, and the Canada First Research Excellence Fund.

\bibliographystyle{ieeetran}
\bibliography{qradar_refs_20190227}

\end{document}